\documentclass{JHEP3}

\usepackage{amssymb}
\usepackage[mathscr]{eucal}
\usepackage{amsmath}
\usepackage{graphicx}
\usepackage{epsfig}
\usepackage{cite}
\usepackage{afterpage}
\usepackage{multirow}

\usepackage{amssymb}

\newcommand{\be}{\begin{eqnarray}}
\newcommand{\ee}{\end{eqnarray}}
\newcommand{\ba}{\left( \begin{array}{ccc}}
\newcommand{\ea} {\end{array} \right)}
\newcommand{\bv}{\left( \begin{array}{c}}
\newcommand{\ev} {\end{array} \right)}

\newcommand{\dd}{\mathrm{d}}

\newcommand{\gsimm}{\raise.3ex\hbox{$>$\kern-.75em\lower1ex\hbox{$\sim$}}}
\newcommand{\lsimm}{\raise.3ex\hbox{$<$\kern-.75em\lower1ex\hbox{$\sim$}}}

\title{Modifying Gravity at Low Redshift}

\author{Philippe Brax \\
  Institut de Physique Th\'eorique, CEA, IPhT, CNRS, URA 2306,
  F-91191Gif/Yvette Cedex, France  \\ E-mail:
  \email{philippe.brax@cea.fr}}

\author{Carsten van de Bruck \\ Department of Applied Mathematics,
  University of Sheffield Hounsfield Road, Sheffield S3 7RH, United
  Kingdom \\ E-mail: \email{c.vandebruck@sheffield.ac.uk}}

\author{Anne-Christine Davis \\ Department of Applied Mathematics and
  Theoretical Physics, Centre for Mathematical Sciences, Cambridge CB3
  0WA, United Kingdom \\ E-mail: \email{a.c.davis@damtp.cam.ac.uk}}

\author{Douglas Shaw \\ Queen Mary University of London, Astronomy Unit,
Mile End Road, London E1 4NS, United Kingdom \\ E-mail: \email{d.shaw@qmul.ac.uk}}

\date{today}

\abstract{We consider the growth of cosmological perturbations in modified gravity models where a scalar field mediates
a non-universal Yukawa force between different matter species. The growth of the density contrast is altered for scales below the Compton wave-length of the scalar field. As the universe expands, the Compton wave-length varies in time in such a way that scales which were outside the range of the scalar field force may  feel it at a lower  redshift. In this case, both the exponent $\gamma$ measuring  the growth of Cold Dark Matter perturbations and the slip function representing  the ratio of the two Newtonian potentials $\psi$ and $\phi$ may differ from their values in General Relativity at low redshift. }

\begin{document}

\section{Introduction}
\label{sec:introduction}
Gravity has been precisely tested in the solar system with tight constraints on fifth forces and violations of the equivalence principle. Its non-linear regime has even been probed in the binary pulsar systems (for a review, see e.g. \cite{Will:2001mx,Uzan:2009ri}). So far no significant deviation from general relativity has been detected. On the other hand, the discovery of cosmic acceleration in 1998 \cite{Perlmutter:1998np,Riess:1998cb} has prompted a rich and renewed bout of activity on models where gravity could be modified (for an overview, see e.g. \cite{Copeland:2006wr} and references therein). This has been triggered by the difficulty of explaining cosmic acceleration with dark energy models. Indeed, dark energy models  with a runaway potential of the Ratra-Peebles type for a real scalar field $\chi$  suffer from the extreme smallness of the mass of $\chi$: today $m_\chi =O(H_0)\approx 10^{-43}$ GeV \cite{Ratra:1987rm}. Such a small mass would leads to $O(1)$ deviations from Newton's law over scales smaller than $\hbar c/ m_{\chi} \sim O({\rm Gpc})$, if the scalar field $\chi$ couples to matter with a strength similar to that of gravity.  Local tests of gravity strongly rule out such deviations over scales larger than $0.1{\rm mm}$ \cite{Will:2001mx}. Generally these constraints are taken to imply that if $\chi$ does couple to baryonic matter it does so with a strength much less than gravity.  This result is valid when the model does not have a chameleon property whereby the mass of the scalar field becomes environment dependent \cite{Khoury:2003rn,Khoury:2003aq,Mota:2006ed,Mota:2006fz}. In particular, chameleon models evade solar system tests thanks to a thin shell effect with a large suppression of the scalar field force for large enough bodies such as the Sun. Of course, it could also be that baryonic matter does not couple to dark energy at all, which would still allow for non-trivial interactions between cold dark matter and  dark energy \cite{Wetterich:1994bg,Anderson:1997un,Amendola:1999er,Farrar:2003uw,Maccio:2003yk,Mainini:2005fe,Mainini:2006zj}. In this case, gravitational experiments in the solar system would not be influenced by the scalar field at all. In fact, the thin shell effect of chameleonic theories implies that this is effectively what happens for structures up to galaxy clusters. In a sense, the coupling of the scalar field has been effaced by the thin shell property and an effective model emerges where large scale fluctuations in dark matter are the only `species' which couple to dark energy. Gravity could also be modified on very large scales as in the DGP model on the accelerating branch (see e.g. \cite{Tanaka:2003zb,Lue:2004rj,Koyama:2005kd} and \cite{Lue:2005ya} for a review). In this case, acceleration is entirely due to the modified structure of gravity.

At the background cosmological level, models of modified gravity such as $f(R)$ theories and its chameleonic siblings are extremely close to a $\Lambda$CDM model. This is also true of large scale modified models of gravity such as DGP. Testing the validity of these different approaches and distinguishing them can be envisaged at the perturbative level
(for an investigation about the growth of perturbations in $f(R)$ models, see e.g. \cite{Gannouji:2008wt,Tsujikawa:2009ku}). Indeed, structure formation is sensitive to fine details in the modification of gravity. In particular, the Newton potential in these models does not remain unique but is embodied in two realizations $\psi$ and $\phi$ with very different roles. First of all, the propagation of light signals is affected by $\psi+\phi$. This is important when considering weak lensing or the ISW effect. On the contrary, peculiar velocities are sensitive to $\psi$ while the baryonic density contrast is related to $\phi$ via the Poisson equation. All in all, depending on which correlation between lensing experiments and either peculiar velocity data or galaxy counts has been analysed, one may find different guises of the modification of gravity. Due to the coupling of the scalar field to matter, the growth of the CDM and baryonic density contrasts is also modified.

Constraints on the coupling of baryons and dark matter to dark energy have been imposed on various scales. The tightest results have been obtained using CMB data for models with a coupling to dark matter only (for a recent analysis, see \cite{Bean:2008ac} and also \cite{Brookfield:2007au,LaVacca:2009yp} for a discussion of coupling dark energy and neutrinos to dark energy). It has been shown that the deviation of the effective Newton constant to the baryonic one cannot exceed 5 percent at scales of order 10 Mpc while the ratio $G_{\rm eff}/G$ could be as large as $2.7$ for scales of 1 Mpc. A tighter bound coming from tidal disruption of dwarf galaxies amounting to a deviation of less than 4 percent at scales of order 100 kpc has been quoted although a comparison with data needs to be performed\cite{kam1,kam2}.
The modification of Newton's constant for models with a coupling $\beta_{DM}$ to dark matter only reads \cite{Amendola:2003wa,Brookfield:2007au}
\begin{equation}
G_{\rm eff}= G \left(1+ \frac{\alpha_{DM}}{1+ (k\lambda_c)^{-2}}\right)
\end{equation}
where $\alpha_{DM} = 2\beta^2_{DM}$, $k$ is the co-moving wave number and $\lambda_c$ the co-moving range of the scalar field force which we will call the Compton length in the following. For scales larger than the Compton length, we have $G_{\rm eff}\approx G$ while well inside the Compton scale we find
$G_{\rm eff}\approx (1+ \alpha_{DM})G$.
CMB data are therefore compatible with a Compton length in between 1 Mpc  and 10 Mpc.

Deviations from General Relativity can be parametrized in a model independent fashion in terms of the slip function $\Sigma$ and the growth rate, $\gamma$, (see e.g. \cite{Amendola:2007rr}).  These are defined by:
\be
\nabla^2(\phi+\psi) &=& 3\Sigma \Omega_{m} H^2 \delta, \\
\gamma &=& \frac{\ln f}{\ln \Omega_{m}}, \ \  \  \ f=\frac{d\ln \delta}{d\ln a}
\ee
where $\delta$ is the density contrast. The definition of $\gamma$ is equivalent to $f = \Omega_{\rm m}^{\gamma}$.  Alternatively one could work in terms of $\eta$ rather than $\Sigma$ where  $\eta=\phi/\psi$. These two definitions are equivalent when $\Sigma$ and $\eta$ are constant and $2\Sigma=1+\eta^{-1}$.  In General Relativity, $\Sigma=\eta = 1$ and $\gamma \approx 0.55$.

In realistic models the Compton length varies in time, as the (effective) mass of the scalar field evolves in time. In some models such as the chameleonic ones, the Compton length decreases with time. It may happen that galactic scales enter the Compton length at a redshift $z^{\ast}$ and therefore the growth of structures for such scales is subsequently affected.  In these models $\Sigma$ and $\gamma$ would deviate  from their GR values only at late-times ($z<z^{\ast}$). Moreover, these parameters become time-dependent. Hence our analysis of  such  late-time modifications of gravity in scalar-tensor theories provide a template for the redshift dependence of $\Sigma$ and $\gamma$. This could be useful for future observational studies of modified gravity.

In the following, we will analyse the growth of structure in linearly coupled models when scales enter the Compton length and gravity is henceforth modified.
The paper is organized as follows: In Section 2 we discuss the perturbation equations and  the evolution of density perturbations. In Section 3 we discuss what happens if a cosmological perturbation crosses the Compton length. We study the growth of perturbations across the jump and find analytical expressions for the perturbation growth. The phenomenological consequences are explained in Section 4. Our conclusions can be found in Section 5.

\section{Cosmological Perturbations in Linearly Coupled Scalar Field Models}
\subsection{Cosmological perturbations}
We are interested in the growth of cosmological perturbations in models where matter interacts with a scalar field. For  convenience we work in the Einstein frame where the gravity equations take the usual form.  This scalar field can mediate a new force between matter species which is non-universal. We are working in the Newtonian conformal gauge in which the metric reads:
\be
\dd s^2 = -a^2(\tau) (1+2\psi(x))\dd\tau^2 + a^2(\tau)(1-2\phi(x))\dd \mathbf{x}^2.
\ee
The Einstein equation
\be
R_{\mu\nu}-\frac{1}{2} R g_{\mu\nu}= 8\pi G T_{\mu\nu}
\ee
has the usual form and the total energy-momentum tensor is:
\be
T_{\mu \nu} &=& \sum_{A} e^{\kappa_4\beta_{A}\chi}\hat{\rho}_{(A)}u_{\mu}^{(A)}u_{\nu}^{(A)}  \\ &&+ \nabla_{\mu}\chi\nabla_{\nu}\chi - \frac{1}{2}g_{\mu \nu}\left[(\nabla \chi)^2 + m^2 \chi^2\right]. \nonumber
\ee
where $A$ labels the different matter species and $u_{\mu}^{(A)}u_{\nu}^{(A)}g^{\mu \nu} = -1$. We have also introduced $\kappa_4= \sqrt{8\pi G}$. Notice the explicit coupling of the scalar field to matter. The couplings $\beta_A$ can be all different from each other.
We assume that the different matter species interact only gravitationally. Matter is conserved  implying that  $\nabla_{\mu}(\hat{\rho}_{(A)}u_{(A)\mu}) = 0$.  Taking the divergence of $T_{\mu \nu}^{(A)}$ and requiring it to vanish  in order to  satisfy the Bianchi identities gives the  $\chi$ field equation:
\be
\nabla^2 \chi = m^2 \chi +\sum_{A}\kappa_4\beta_{A} \hat{\rho}_{(A)}e^{\kappa_4\beta_{A}\chi}.
\ee
The acceleration of test particles $a^{(A)}_{\mu}= u^{(A)\nu} \nabla_\nu u_{\mu}^{(A)}$ is influenced by the presence of the scalar field and reads
\be
a^{(A)}_{\mu} = -\kappa_4\beta_{A} D^{(A)}_{\mu} \chi, \label{accEqn}
\ee
where $D^{(A)}_{\mu} = \nabla_{\mu} + u^{(A)}_{\mu} u^{(A)\nu}\nabla_{\nu}$.  We also define for convenience $\rho_{(A)} = e^{\kappa_4\beta_{A}\chi}\hat{\rho}_{(A)}$.
Since we are in the Einstein frame, the  matter density $\rho_{(A)}$ is the Einstein frame matter density which is not generally conserved.

We define $u^{(A)i} = v_{(A)}^i/a$ and then take $u_{0}^{(A)} > 0$ and assume $\vert \mathbf{v}_{(A)}\vert^2 \ll 1$ and $\psi, \phi \ll 1$ and  $\epsilon \ll 1$ so that to leading order in each term:
\be
u^{(A)\mu} \approx a^{-1}(1-\psi + \frac{1}{2}\mathbf{v}_{(A)}^2, v^{i}_{(A)})^{\rm T}
\ee
Hence, including all potentially leading order terms, we have the Euler equation from Eq. (\ref{accEqn}):
\be
a^{-1} \dot{v}_{(A)}^{i} + H v^{i}_{(A)} + a^{-1} v^{j}_{(A)}\partial_{j}v^{i}_{(A)} \approx -a^{-1}\left[ \psi+ \beta_{A}\chi\right]_{,i}.
\ee
where $H = \dot{a}/a^2$.  Notice that matter feels $\psi + \beta_A \chi$ and not the Newton potential $\psi$ only.

We also define $\hat{\rho}_{A} = \rho^{(0)}_{A} e^{\delta_{A}}/ a^3 $, and $\rho_{A}^{(0)} = {\rm const}$.  Then keeping all potentially leading order terms, the conservation equation reads:
\be
a^{-1}\dot{\delta}_{A} + a^{-1}v^{i}_{(A)}\delta_{A,i} \approx - a^{-1}\partial_{i}v^{i}_{(A)}
\ee
We define $\mathcal{H} = aH$ and then consider perturbations to linear order. Using the divergence of the velocity field $\theta_{A} = v_{(A),i}^{i}$ we have:
\be
\dot{\delta}_{A} &\approx& -\theta_{A}-3\dot{\phi}, \\
\dot{\theta}_{A} + \mathcal{H}\theta_{A}&\approx& -\left[ \psi+ \beta_{A}\chi\right]_{,ii} \equiv -\psi_{A,ii}. \label{eqn:pecvel}
\ee
Now to leading order on sub-horizon scales we have, linearizing the matter perturbations:
\be
\phi_{,ii} &\approx& 4\pi G a^2 \sum_{A} \bar{\rho}_{(A)}\delta_{A}, \\
\phi &\approx & \psi, \\
\chi_{,ii} &\approx& a^2m^2 \chi +  a^2 \sum_{A} \kappa_4\beta_{A}\bar{\rho}_{(A)}\delta_{A} \label{eqn:matterperturbations}
\ee
where the last one holds because we assume the energy density of the scalar field perturbation to be small. We have therefore:
\be
\ddot{\delta}_{A} + \mathcal{H}\dot{\delta}_{A} &\approx& \psi_{A,ii} -3\ddot{\phi}-3\mathcal{H}\dot{\phi}.
\ee
On sub-horizon scales the last-terms can be ignored. Performing a Fourier transform we arrive
at \cite{Amendola:2003wa,Brax:2004qh, Brookfield:2007au}:
\be
\ddot{\delta}_{A} &+& \mathcal{H}\dot{\delta}_{A} \approx -k^2\psi_{A} \\ &=& \frac{3}{2}\mathcal{H}^2 \sum_{B} \Omega_{B}(a) \delta_{B}(1+\alpha_{AB}(am/k)), \nonumber
\ee
where
$$
\alpha_{AB}(x) = \frac{2\beta_{A}\beta_{B}}{1+x^2}.
$$
and $x= am/k$.
The modification of gravity is sensitive to the Compton wave-length $\lambda=1/m$. Typically when the co-moving wave-number $k$ becomes larger than the inverse Compton length $\lambda_c^{-1}= ma$, the growth of structures is affected\cite{green}.

Typical models with this behaviour have been constructed. Originally, the mass varying dark matter model of \cite{Anderson:1997un} is such that only dark matter has a coupling to $\chi$. Moreover, the mass $m$ increases with time (see also \cite{Amendola:2007yx} for a model with growing neutrino mass). As a result for a given co-moving scale $L=k^{-1}$, gravity would have been modified in the past and come back to normal once $L\ll \lambda_c$. This behaviour is also characteristic of models with a constant mass $m$. On the other hand, chameleon models\cite{Khoury:2003rn,Brax:2004qh,Mota:2006ed} are such that all species couple to the scalar field at the linear level. At the non-linear level, a thin shell appears which suppresses the coupling of $\chi$ to baryons. Phenomenologically, this can be implemented in this setting by imposing that only cold dark matter couples to $\chi$ for scales below the size of clusters. The Compton length can increase or decrease in time. Models where it decreases can be obtained with an inverse power law potential of index $n$ for which $ma\sim a^{-(n+4)/(n+2)}$ decreases with time. In this case, gravity is modified at late time when scales become within the Compton length.

\subsection{Growth of structures: $\gamma_A$ and $\gamma_B$}

In this section we focus on a two species system   $A$ and $B$, and  $\Omega_{B} \gg \Omega_{A}$ so the matter density is then $\Omega_{\rm m} \approx \Omega_{\rm B}$.  In practise $B$ would be Cold Dark Matter (CDM)  while $A$ would be the baryons.  The evolution of structures is governed by
\be
\ddot{\delta}_{A} &+& \mathcal{H}\dot{\delta}_{A} \approx \frac{3}{2}\mathcal{H}^2 \Omega_{B}(a) \delta_{B}(1+\alpha_{AB}), \\
\ddot{\delta}_{B} &+& \mathcal{H}\dot{\delta}_{B} \approx \frac{3}{2}\mathcal{H}^2 \Omega_{B}(a) \delta_{B}(1+\alpha_{BB}).
\ee
We define $\alpha(x) \equiv \alpha_{BB}(x)$ and
$$
1+\xi(x)  = \frac{1+\alpha_{AB}(x)}{1+\alpha_{BB}(x)}.
$$
When $x \gg 1$ or $x \ll 1$, $\xi$ is a constant. Hence whenever  the  scales are either well within, or well beyond the Compton length, $\xi \approx {\rm const}$ and it is clear then that
\be
\hat{\delta}_{A} = \frac{\delta_{A}}{1+\xi} + {\rm const},
\ee
satisfies the same evolution equation as $\delta_{B}$.   Provided the scales of interest do not cross the Compton length (so $\xi \approx {\rm const}$), at late-times we can assume $\hat{\delta}_{A} = \delta_{B}$ so $\delta_{A} = (1+\xi)\delta_{B} + \Delta_{0}$; $\Delta_0$ is a constant. At late times then $\delta_A \sim (1+\xi)\delta_B$.   More generally we can define an effective value of $\xi$ thus:
\be
\delta_A = (1+\xi_{\rm eff})\delta_{B} +\Delta_0. \label{xi_eff}
\ee
It follows that the bias, $b_{AB} =\delta_{A}/\delta_{B}$, between species A is given by:
\be
b_{AB} = (1+\xi_{\rm eff})  + \frac{\Delta_0}{\delta_{B}}.
\ee
At late-times $\Delta_0/\delta_B \rightarrow 0$ and so the dominant contribution comes from $1+\xi_{\rm eff}$.  We note that if $\alpha_{AB}=\alpha_{BB}$ then $\xi \equiv 0$ and $\xi_{\rm eff} = 0$.  In this case, $\delta_A = \delta_B + \Delta_0$ which is precisely the same equation as one encounters in GR.

We suppose that the background cosmology is well approximated by the $\Lambda$CDM evolution; in chameleon models, this is realized when  $m^2 \gg H^2$.  This implies that $\bar{\rho}_{\rm B}, \bar{\rho}_{\rm A} \propto a^{-3}$ and furthermore since $\Omega_{\rm m} \approx \Omega_{\rm B} \gg \Omega_{\rm  A}$:
\be
\Omega_{\rm B} \approx \Omega_{\rm m} = \frac{1}{1+ a^3 (1-\Omega_{\rm m0})/\Omega_{\rm m0}},
\ee
where $\Omega_{\rm m0}$ is the value of $\Omega_{\rm m0}$ at $a=1$.

We define:
\be
f_{B}(\ln a, k) = \frac{\dd \ln \delta_{B}}{\dd \ln a},
\ee
and find that this satisfies
\be
\left[2 -\frac{3\Omega_{\rm B}(a)}{2}\right]f_{B} + f^2_{B} +f^{\prime}_{B} \approx \frac{3}{2}\Omega_{\rm B}(a)(1+\alpha_{\rm BB}). \label{fBeqn}
\ee
where ${}^{\prime}=d/d\ln a$. Henceforth $\Omega_{\rm A}$ does not enter the field equation and we replace $\Omega_{\rm B}$ by $\Omega_{\rm m}$.

We now assume that we know the solution, $f_{\rm GR}$, of Eq. (\ref{fBeqn}) when $\alpha_{\rm BB} \equiv 0$ i.e. in General Relativity. We parametrize the solution in the modified gravity scenarios ($\alpha_{BB} \neq 0$) thus: $f(\ln a, k) = (1+g_B(\ln a)) f_{\rm GR}(\ln a)$. This gives:
\be
(1+g_B)g_B + \frac{3 \Omega_{\rm m}}{2 f_{\rm GR}^2}g_B + g^{\prime}_B f_{\rm GR}^{-1} = \frac{3\alpha_{\rm BB} \Omega_{\rm m}}{2f_{\rm GR}^2}. \label{gdef}
\ee
In the $\Lambda$CDM model it was found that $f_{\rm GR} \approx \Omega_{\rm m}^{0.55}$ \cite{Linder:2005in}. Since $\Omega_{\rm m} \approx \Omega_{\rm B}$, $\Omega_{\rm B} / f_{\rm GR}^2 \approx \Omega_{\rm m}^{-0.1}$, which varies  only very slowly up to the present epoch (a different approximation was be found in \cite{Lahav:1991wc}, but for the purpose of this paper, we use the one found in \cite{Linder:2005in}). We find the analytical solution to Eq. (\ref{gdef}) under the approximation $\Omega_{\rm m}/f_{\rm GR}^2 \approx 1$ i.e. $f_{\rm GR} \sim \Omega_m^{0.5}$. For the time being, we also approximate $\alpha_{\rm BB}$ by a constant. We then have:
\be
g_B = g(\alpha_{\rm BB}) \approx -\frac{5}{4} + \sqrt{\frac{25}{16} + \frac{3}{2}\alpha_{\rm BB}}. \label{gEqn}
\ee
Thus with $f_{\rm GR} = \Omega_{\rm m}^{\gamma_{\rm GR} \approx 0.55}$ we have $f = \Omega_{\rm m}^{\gamma_{\rm B}}$ where:
\be
\gamma_{\rm B}(a,k) - \gamma_{\rm GR} \approx \frac{\ln(1+g_B(a,k))}{\ln \Omega_{\rm m}}.
\ee
We could similarly define $g_A$ and $\gamma_{A}$ to describe the evolution of the $A$-type matter species. Under the assumption that couplings are constant, we have $\delta_{\rm A} = (1+\xi)\delta_{\rm B} + \Delta_0$ and hence:
\be
(1+g_{\rm A}) =  (1+g_{\rm B})\left[\frac{1+\xi}{b_{AB}}\right]. \label{gAeqn0}
\ee
At late-times the term in square brackets tends to $1$ so $g_{\rm A} \rightarrow g_{\rm B}$.

However, as we shall see, if the Compton wavelength is crossed and $\xi \neq 0$ (i.e $\beta_{\rm A}\neq \beta_{\rm B}$) then the simple correspondence of Eq. (\ref{gAeqn0}) is broken and it is entirely feasible that  $\gamma_{B}$ could deviate greatly from $\gamma_{\rm GR}$ whilst $\gamma_{A}$ hardly changes at all.

In general though, we would have $\gamma_A, \,\gamma_{B} \neq \gamma_{\rm GR}$ if $\alpha_{\rm BB} \neq 0$. We study the effect of the jump in $\alpha_{\rm BB}$ and $\xi$ when scales get inside the Compton length in the following section.
\subsection{The Slip Functions}
Being interested in the modification of gravity, we now focus on the two gravitational potentials, $\phi$ and $\psi$ and their ratio described by the $\eta = \phi/\psi$ parameter or $2\Sigma = 1+\eta^{-1}$.

The quantity $\psi+\phi$ is invariant under a conformal rescaling of the metric, however, individually, $\phi$ and $\psi$  are not.  A choice of conformal frame is essentially a choice of standard ruler and clock.  In General Relativity there is a preferred and obvious choice of conformal frame where the Newtonian constant, $G$, is fixed and the energy momentum tensor is conserved ensuring that the masses of particles are constant.  In modified gravity such a frame choice is not generally possible.  In some theories (i.e. those with a universal coupling to a scalar field), there exists what is commonly known as the Jordan frame.   In this frame the matter energy momentum tensor is conserved, particle masses are constant and non-gravitational physics is independent of space-time position. It is for this reason that the Jordan frame is often referred to as the "physical frame", however this nomenclature can be misleading. Additionally in the Jordan frame the effective Newton  constant varies with space and time. In scalar-tensor theories, including those with multiple scalar fields and different couplings, one may always define an Einstein frame where $G$ is constant, but $T^{\mu \nu}_{\rm m;\mu} \neq 0$. In this frame the particle masses depend on space time positions and hence so does local non-gravitational physics. However, this does not mean that the Einstein frame is in some sense `unphysical'.  Both Jordan and Einstein frames, and indeed any other choice of conformal frame, are physical in the sense that provided one does not assume or require quantities to be constant that are not and interprets all measured quantities correctly, they represent a perfectly accurate descriptions of nature.

\subsubsection{Weak Lensing Measurements}

We first consider weak gravitational lensing measurements, which are sensitive to $\phi+\psi$ (for a discussion of weak lensing in scalar tensor theories, see e.g. \cite{Schimd:2004nq}).  In the Einstein frame $\nabla^2 \phi = 4\pi G \delta \rho$, if one assumes that this is generally true and takes $\psi = \eta^{-1}\phi$ where $\eta$ is assumed to be constant, we have:
\be
\nabla^2 (\phi+\psi) = 8\pi G \Sigma \delta \rho, \label{SigmaKapDef}
\ee
where $2\Sigma = 1+\eta^{-1}$. In modified gravity models such as the ones we are considering, the parameter $\Sigma$ defined by (\ref{SigmaKapDef}) is not constant. In general,
we will take this form of the Poisson equation in the Einstein frame as the definition of $\Sigma$. In General Relativity, we have $\Sigma=1$. In the type of scalar-tensor theory that we are considering, the absence of anisotropic stress in the Einstein frame implies that $\Sigma\equiv 1$. Let us consider another frame obtained by a Weyl rescaling of the Einstein metric; $g_{\mu\nu}\to \tilde g_{\mu\nu}= e^{-2w} g_{\mu\nu}$ where $w\ll 1$ is an arbitrary function. Under this change of metric, the Newton potentials are transformed as $\psi\to \psi+w$ and $\phi\to \phi-w$, implying that $\phi+\psi$ is frame invariant. This is not the case of the energy density of non relativistic matter $\rho=-g^{\mu\nu}T_{\mu\nu}$ which transforms as $\rho \to \tilde \rho= e^{4w} \rho$ while $\tilde G= e^{-2w} G$ . Using $\nabla^2 = e^{-2w}\tilde \nabla^2$ and the fact that for a given function $w$ which is not a field itself $\delta \rho= e^{-4w} \delta\tilde \rho$, we find that $\Sigma$ is invariant under a Weyl rescaling of the metric. This is a major advantage of defining $\Sigma$ using the modified Poisson equation. Notice  that when the Weyl transformation is designed to efface the coupling of a particular species A to $\chi$, one must choose a field dependent
\begin{equation}
w(\chi)= -\kappa_4\beta_A \chi,
\end{equation}
as it is the case for the theories considered here. The resulting metric $\tilde g_{\mu\nu}$ is the Jordan metric for species $A$. Unless the couplings are universal, this frame is not the Jordan frame for species $B\ne A$. In this frame we have $\delta \rho= e^{4\kappa_4\beta_A \chi} (\delta \tilde \rho +4 \kappa_4\beta_A \delta \chi \tilde \rho)$. Hence $\Sigma$ is frame independent in this case if and only if $ \kappa_4 \beta_A \vert \delta \chi\vert \ll \frac{\delta \tilde \rho}{\tilde \rho}$,, i.e. if the scalar field fluctuations can be neglected. When the coupling is universal and upon using (\ref{eqn:matterperturbations}), this is always true as long as $\beta^2 \Omega_{\rm CDM} H^2 \ll m^2 \frac{\delta\rho_{\rm CDM}}{\rho_{\rm CDM}}$. This gives a bound on $\beta$ depending on the ratio $m/H\gg 1$.  When this bound is satisfied,  the slip parameter is frame independent.

We denote $\Sigma$ defined by Eq. (\ref{SigmaKapDef}) by  $\Sigma_{\kappa}$.  Provided one can measure $\delta \rho$, $\Sigma_{\kappa}$ can be extracted from weak-lensing measurements.
Fourier-transforming Eq. (\ref{SigmaKapDef}) gives:
\be
-k^2 (\phi+\psi) &=& 3\mathcal{H}^2 \Sigma_\kappa \sum_{C} \Omega_{\rm C}\delta_C  \approx 3\mathcal{H}^2  \Omega_{\rm m}\Sigma_\kappa \delta_{B}, \label{SigmaKapDef2} \\
&=& 3\mathcal{H}^2 \Omega_{\rm m}  \left[\Sigma_\kappa  b_{AB}^{-1}\right] \delta_{A}. \nonumber
\ee
where we have used the definition of the bias function in the last equality.
Using weak-lensing measurements alone one can compare the ratio of $\Sigma^2 \delta_{B}^2$ with its value at an earlier time, which provides a measurement of the growth rate of $\Sigma_{\kappa} \delta_B$.  We define
\begin{equation}
\delta_{B}(z,k) = D_{B}(z,k)\delta_{i}(k),
\end{equation}
so $\dd \ln D_{B} /\dd \ln a  = f_{B}$
and $\delta_i$ is the primordial value of the perturbation. The initial $\delta_{i}^2(k)$ is proportional to the primordial power spectrum and is measured by CMB experiments such as WMAP.  Therefore by combining weak-lensing measurements with an ansatz or measurement of the primordial power-spectrum one measures not $\Sigma_\kappa(z,k)$ but the combination:
\be
\Sigma_{\kappa \kappa} = \Sigma_{\kappa}(z,k) D_{B}(z,k).
\ee
Another method of extracting $\Sigma_\kappa$ would be to directly measure $\delta_A$. For instance if $A$ represents galaxies this could be done using galaxy surveys.  From cross-correlation of weak-lensing and $\delta_A$ one can then extract the quantity:
\be
\Sigma_{\kappa A} = \Sigma_{\kappa}(z,k) b_{AB}^{-1}(z,k)
\ee
Combining measurements of $\delta_A$ with a measurement of the primordial power spectrum provides:
\be
D_{A}(z,k) = b_{AB}(z,k)D_B(z,k).
\ee
We note that $D_B$ and $D_A$ are determined by $\gamma_B$ and $\gamma_A$ respectively and by the cosmological model through $\Omega_{m}(a)$.  If one assumes a GR growth rate for species $B$ then the measured value of $\Sigma_{\kappa}$ is $\Sigma_{\kappa m}$:
\be
\Sigma_{\kappa m} = \Sigma_{\kappa} \frac{D_B}{D_{\rm GR}}.
\ee
Alternatively if one assumes a particular ansatz for $b_{AB} = \bar{b}(z,k)$ then one measures:
\be
\Sigma_{\kappa b} = \Sigma_{\kappa} \frac{\bar{b}}{b_{AB}}.
\ee
In a scalar-tensor theory, such as the class of theory considered here, $\Sigma_{\kappa} =1$. However since one does not measure $\Sigma_\kappa$ on its own, the measured value of $\Sigma_{\kappa}$ will depend on the ansatz one makes for either the growth rate $\gamma_B$ or bias $b_{AB}$ as well as the cosmological model, for instance:
$$
\Sigma_{\kappa m} = \frac{D_B}{D_{\rm GR}}, \qquad \Sigma_{\kappa b} =  \frac{\bar{b}}{b_{AB}}.
$$
If one estimates $\bar{b}$ by fitting to the observables and assuming a GR growth rate $D_A = b_{AB} D_B=\bar b D_{\rm GR} $  would give: $\bar{b}/b_{\rm GR} = D_{\rm B}/D_{\rm GR} $ and $\Sigma_{\kappa m} = \Sigma_{\kappa b}$.

Given that weak-lensing measurements directly probe $\psi+\phi$, one could measure $\eta= \phi/\psi$ directly if one could measure either $\phi$ or $\psi$ directly.  In this case, the frame in which one effectively measures $\eta$ depends on the method one uses to measure $\phi$ or $\psi$.
The most direct method for measuring $\psi$, which does not depend on the bias between different species, is to use  peculiar velocities.    From Eqs. (\ref{eqn:pecvel}) and (\ref{eqn:matterperturbations}) it is clear that the peculiar velocities of species $A$ depend on
$$\psi_A  \approx -(1+\alpha_{AB})\frac{3\Omega_{m}\mathcal{H}^2}{k^2} \delta_{B} = \frac{1}{2}(1+\alpha_{AB})\left[\phi+\psi\right].$$
If $\beta_{A}=\beta_B$ so $\alpha_{AB} = \alpha_{BB}=\alpha$, then $\psi_{A}$ is precisely the value of $\psi$ one would calculate in the Jordan frame.   Similarly if $\beta_A=0$, then $\psi_A$ is equal to the Einstein frame value of $\psi$. Assuming that $\psi\equiv\psi_A$ one would therefore estimate $\eta = \eta_{\theta}$ where:
\be
1+\eta_{\theta} = \frac{\psi+\phi}{\psi_A} = \frac{1-\alpha_{AB}}{1+\alpha_{AB}}.
\ee
We may also define $2\Sigma_{\theta} = 1+\eta_{\theta}^{-1} = 1/(1-\alpha_{AB})$. In any scenario where species $A$ does not couple to the scalar field, $\beta_A = 0$, $\Sigma_{\theta} = \eta_{\theta}=1$.  In theories such as $f(R)$ theories, galaxies are expected to be effectively decoupled from the scalar force as the result of the chameleon mechanism which allows for compatibility of the theory with local tests of gravity. In such theories then measurements of the peculiar velocities of galaxies would find $\Sigma = \eta =1$ and hence not reveal any modification of gravity even though large-scale CDM perturbations might well feel an non-negligible fifth force.

\subsubsection{ISW Measurements}
The ISW effect  depends on the quantity $\phi^{\prime}+\psi^{\prime}$.  Given $\Sigma_{\kappa}$ as defined Eq. (\ref{SigmaKapDef2}) is follows that:
\be
-k^2\left[\phi^{\prime} + \psi^{\prime}\right] \approx 3\Omega_{m}\mathcal{H}^2  \Sigma_{I} \left[f_{\rm GR}-1\right] \delta_{B} =  3\Omega_{m}\mathcal{H}^2  \Sigma_{I} b_{AB}^{-1} \left[f_{\rm GR}-1\right] \delta_{A},
\ee
where
\be
\Sigma_{I} = \Sigma_{\kappa}\left[\frac{1-f_B}{1-f_{\rm GR}} + \frac{\Sigma_{\kappa}^{\prime}}{\Sigma_{\kappa}}\right].
\ee
We note that in the simple case where $f_{B}=f_{\rm GR}$ and $\Sigma_{\kappa} = {\rm const}$, $\Sigma_{I}=\Sigma_{\kappa}$.  As with weak-lensing measurements, in practice one would not measure $\Sigma_{I}$ directly but either:
\be
\Sigma_{II} = \Sigma_{I} D_{B},\nonumber
\ee
or
\be
\Sigma_{IA} = \Sigma_{I} b_{AB}^{-1}. \nonumber
\ee
If one assumes a GR growth rate, then the measured of $\Sigma_{I}$ would be:
\be
\Sigma_{Im} = \Sigma_{I} \frac{D_{\rm B}}{D_{\rm GR}}.
\ee
In the class of scalar-tensor theories we have considered $\Sigma_{\kappa} = 1$ and so:
\be
\Sigma_{Im} = \frac{(1-f_{B})D_{\rm B}}{(1-f_{\rm GR})D_{\rm GR}}
\ee

\subsubsection{Summary}
 We have introduced  the slip parameter $\Sigma_\kappa$ using a modified version of the Poisson equation.  Extracting the slip function from data requires an assumption on the growth factor.   Using weak-lensing and assuming a GR growth rate one would measure:
\be
\Sigma_{\kappa m} = \frac{D_{\rm B}}{D_{\rm GR}}, \nonumber
\ee
Using ISW measurements, and under the same assumptions one would extract:
\be
\Sigma_{I m} = \frac{(1-f_{B})D_{\rm B}}{(1-f_{\rm GR})D_{\rm GR}}. \nonumber
\ee
Both expressions are determined by $D_{\rm B}$ which itself is given by $f_{\rm B} = (1+g_B)f_{\rm GR}$ and the growth rate $\gamma_{\rm B}$.  The bias between species $A$ and $B$ depends on $\xi_{\rm eff}$ and $g_{\rm B}$ as does the growth rate of species $A$; $\gamma_{A}$.  In these next section we will find very good analytical approximations to the evolution of $g_{\rm B}$ and $\xi_{\rm eff}$ when the couplings $\beta_{A}$ and $\beta_{B}$ jump at $z=z^{\ast}$.

\section{Effect of a Change in the Coupling}
We now consider what happens when the values of $\alpha_{\rm AB}$ and $\alpha_{\rm BB}$ change at some redshift $z=z^{\ast}$ for instance between a perturbation crosses the Compton wavelength of the scalar field $\chi$.
We first consider the growth function for the dominant matter species $B$ before considering the sub-dominant species $A$.

\subsection{Dominant Species Growth Function, $f_B$}
To simplify the problem, we assume that the coupling function $\alpha_{BB}$ changes abruptly (jumps) at a redshift $z_{\ast}$, i.e. we assume that for $z > z^{\ast}$, $\alpha_{BB} = \alpha_{0}$, and for $z < z^{\ast}$, $\alpha_{BB} = \alpha$ for some $\alpha$ and $\alpha_0$.   We define $g_{0} =g(\alpha_0)$ and $g_{\alpha}=g(\alpha)$. For $z > z_{\ast}$, $\alpha_{BB}$ is taken to be constant back to the far past, and so $g_B = g_{0} = {\rm const}$.  For $z < z_{\ast}$, $g_{B}$  must satisfy Eq.  (\ref{gdef}):
\be
(1+g_B)g_B + \frac{3\Omega_m}{2f_{\rm GR}^2}g_{B} + \frac{g_{B}^{\prime}}{f_{\rm GR}} = \frac{3\alpha \Omega_m}{f_{\rm GR}^2}
\ee
We define $g_{B} = g_{\alpha} + \Delta g$ and using Eqs. (\ref{gdef}) and (\ref{gEqn}) arrive at:
\be
\frac{(\Delta g)^{\prime}}{f_{\rm GR}} = -\left[\frac{3\Omega_m}{2f_{\rm GR}^2} + 1 + 2g_{\alpha} + \Delta g\right]\Delta g.
\ee
Now $f_{\rm GR} = \Omega_m^{\gamma_{\rm GR}}\approx \Omega_m^{0.55}$. As in the derivation of $g(\alpha)$, we take $\Omega/f_{\rm GR}^2 \approx \Omega_{m}^{-0.1}$ to be unity (and constant).  This is equivalent to taking $f_{\rm GR} \approx f_0 = \sqrt{\Omega_{m}}$ for the purposes of finding $g_B$,  $\xi_{\rm eff}$ and $g_A$.   Clearly this is exact when $\Omega_m = 1 $ and is a good approximation generally when $\Omega_m^{-0.1}-1 \ll 1$. Notice that $f_{\rm GR} \approx f_0 =\Omega_m^{1/2}$ is the \emph{only} approximation that we make in what follows.
\FIGURE{\includegraphics[width=7cm]{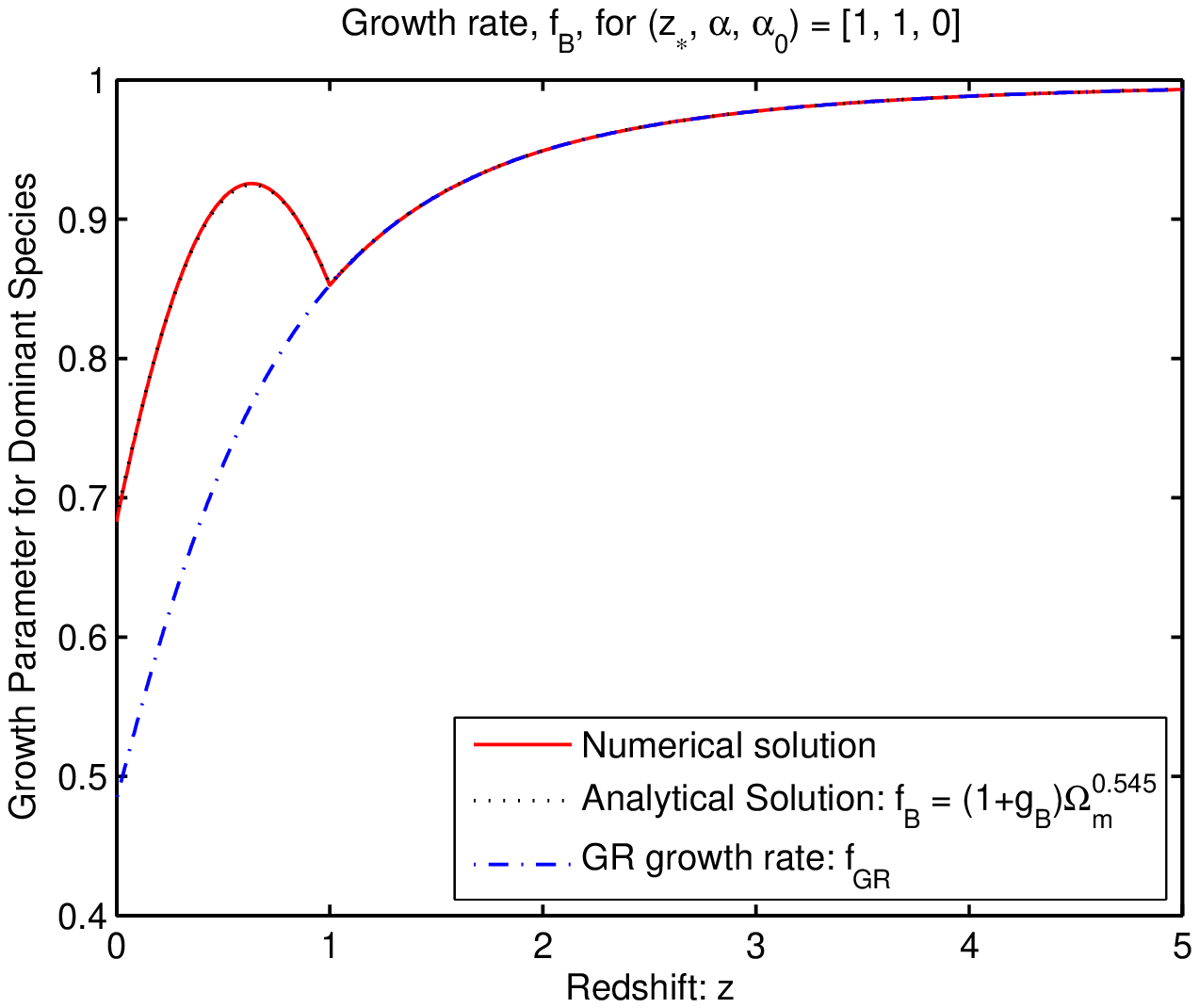}
\includegraphics[width=7cm]{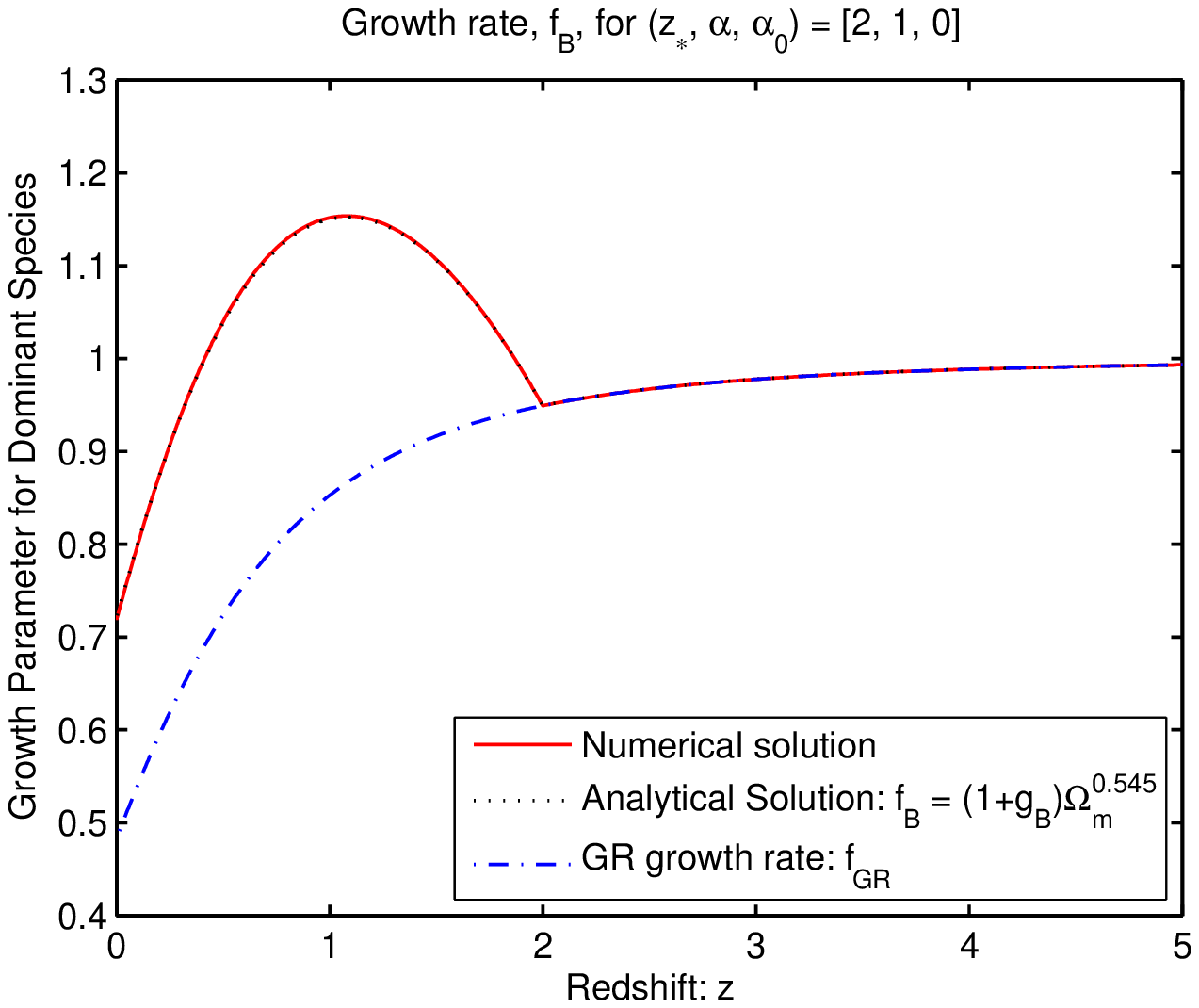}
\includegraphics[width=7cm]{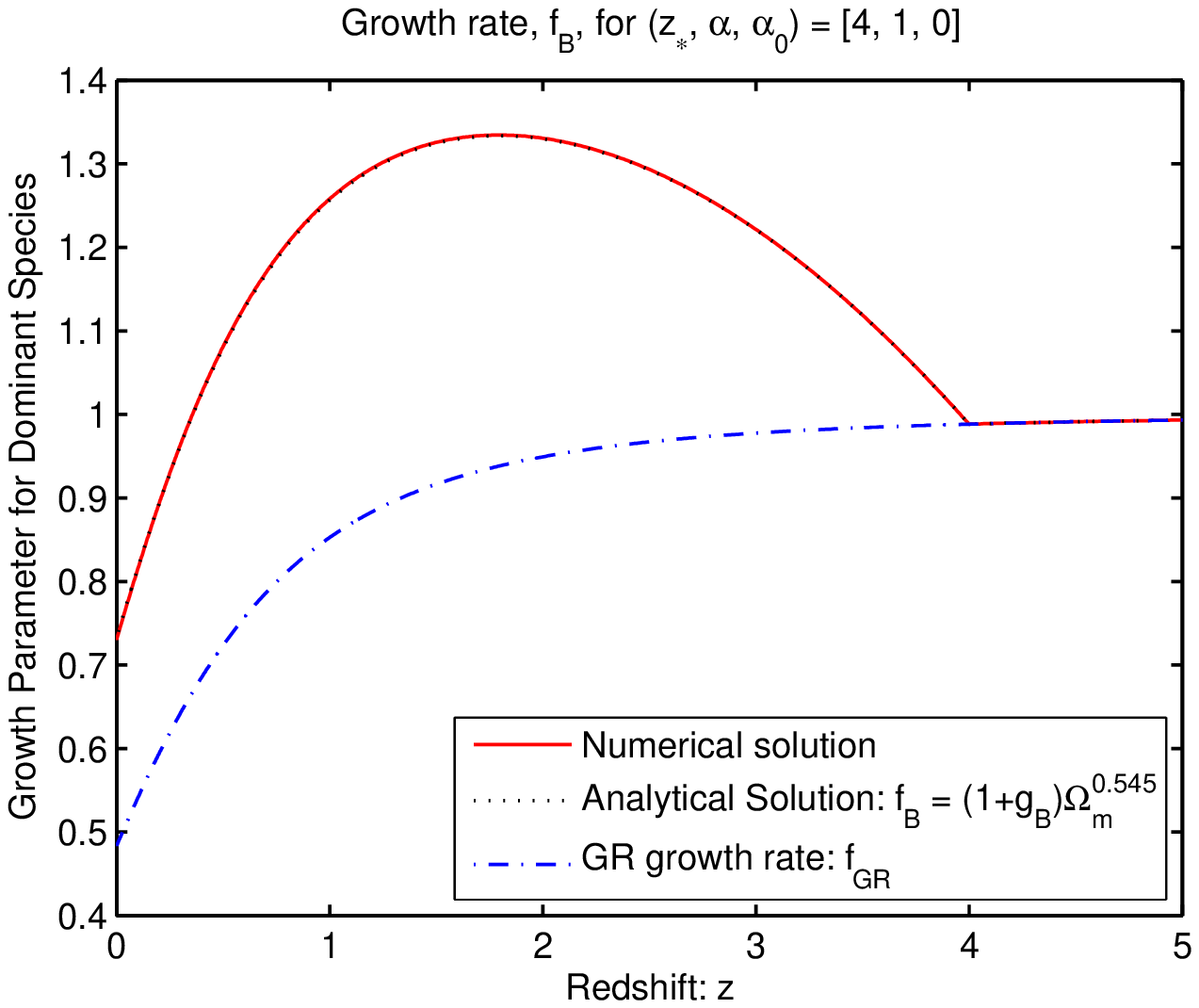}
\includegraphics[width=7cm]{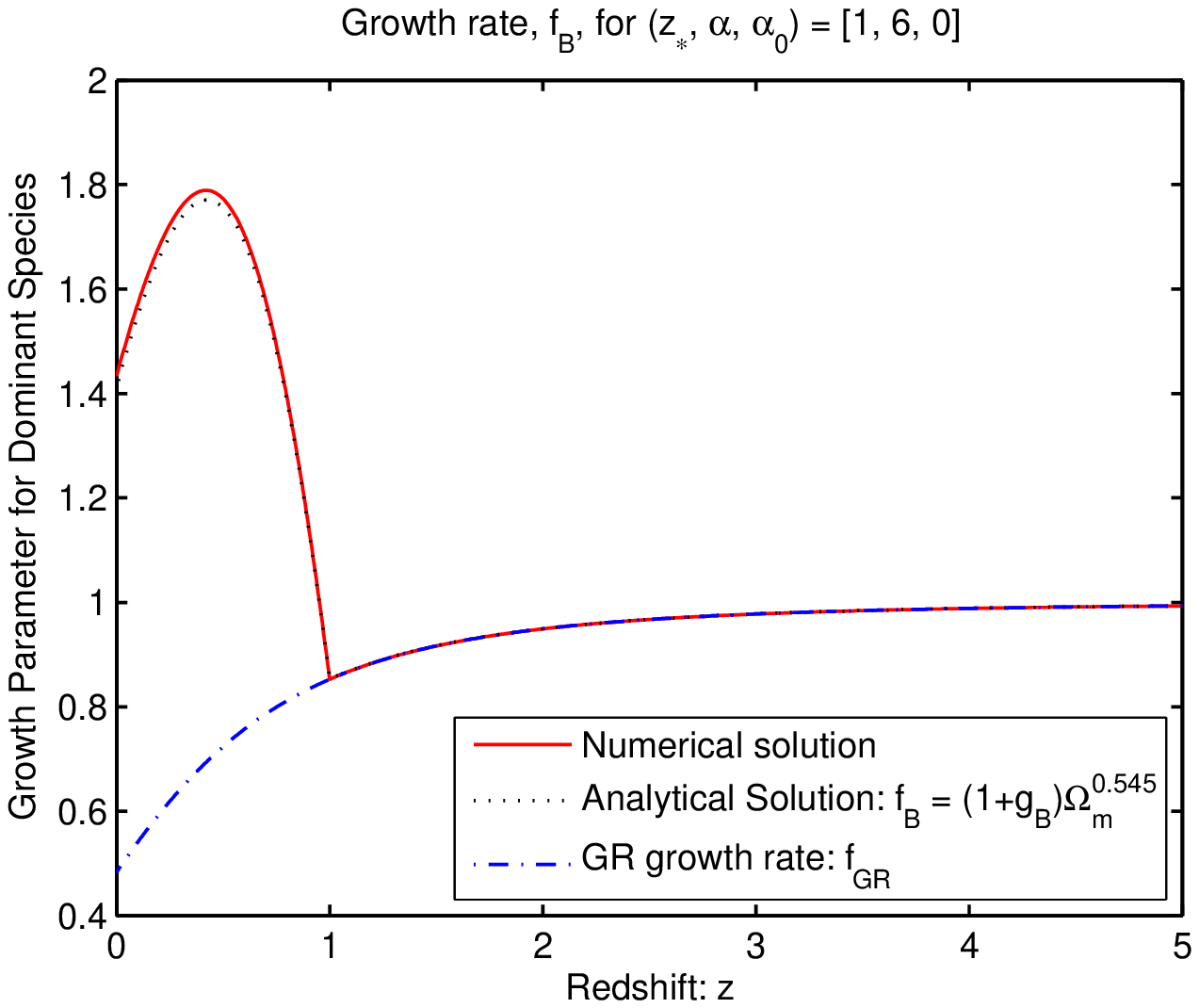}
\caption[]{Sample behaviours for the linear growth rate, $f_B$, of the dominant matter species i.e. dark matter. In all cases shown above the $\alpha_0$, fifth force coupling for $z>z_{\ast}$, is taken to vanish and the late-time coupling $\alpha$, for $z < z_{\ast}$, is a constant.  The plots above show the behaviour for $(z_{\ast},\alpha) = (1,1)$ (top-left), $(2,1)$ (top-right), $(4,1)$ (bottom-left) and $(1,6)$ (bottom-right).   We see that in all cases there is a pronounced deviation from General Relativity where $f_B = f_{\rm GR} \approx \Omega_{m}^{0.545}$ (shown as the dot-dashed blue line).  The solid red line is the exact numerical solution, and the dotted black line is our analytical approximation: $f_B = (1+g_B)\Omega_{m}^{0.545}$. We note that the analytical approximation derived in this work is almost exact at all times.}
\label{FIGfB}}

With this approximation we find that we can further set  $3\Omega_m/2f_0^2 +1 + 2g_{\alpha} \approx 5/2+2g_{\alpha}$ is a constant and we define it to be $\mu_{0}$.  By integration we find:
\be
\frac{\Delta g}{\mu_{0} +  \Delta g} \approx -A_0 D_{0}^{-\mu_0},
\ee
where $A_0$ is a constant of integration and $D_0$ is defined by
$$
D_{0}^{\prime} = f_0 D_{0}
$$
and at $D_0(z_{\ast}) = D_{\ast}$. We have
\be
D_0 = D_{\ast} \frac{K(\Omega_m)}{K(\Omega_m^{\ast})}, \qquad K(\Omega_m) \approx  \left(\frac{1-\sqrt{\Omega_m}}{1+\sqrt{\Omega_m}}\right)^{1/3}.
\ee
In this equation we have defined $\Omega_m^{\ast} = \Omega_m(z=z_{\ast})$.  As $\Omega_m,\Omega_m^{\ast} \rightarrow 1$, $D_0/D_{\ast} \approx a/a_{\ast} = (1+z_{\ast})/(1+z)$.

Now at $z=z_{\ast}$, $g_{B} = g_0$ and so $\Delta g = (g_{0}-g_{\alpha})$ and, hence
\be
A_{0} = \frac{g_{\alpha} -g_{0}}{\mu_0-(g_{\alpha} -g_{0})}.
\ee
Thus we find
\be
g_{\rm B} \approx g_{\alpha} -\frac{\mu_0A_0 D_{\ast}^{\mu_0} }{D_{0}^{\mu_0}+A_0 D_{\ast}^{\mu_0} }, \\
\Rightarrow g_{\rm B} \approx g_{0} + (g_{\alpha}-g_0) G_{B}(D_{0}/D_{\ast};\mu_0,A_0),\nonumber
\ee
where for $X<1$, $G_{B}(X;\mu_0,A_0) = 0$ and for $X>1$:
$$
G_{B}(X>1;\mu_0,A_0) = \left[\frac{X^{\mu_0}-1}{X^{\mu_0} + A_0}\right].
$$
Now:
\be
\left[\ln D_{\rm B}\right]^{\prime} = (1+g_{\rm B})\left[\ln D_{\rm GR}\right]^{\prime}. \nonumber
\ee
Thus it follows that using $D_{\rm GR} \approx D_{0}$ that:
\be
\Sigma_{\kappa m} &\approx&  D_{0}^{g_0} F_{B}(D_{0}/D_{\ast}; \mu_0, A_0), \label{SigKapMEqn}\\
\ee
where  we assume that at some initial time, $z=z_{i}$, $\delta =\delta_{i}$ and $D_0=D_{\rm B} = 1$ while  for $X<1$, $F_{B}(X;\mu_0,A_0)  = 1$.  For $X > 1$.
\be
F_{B}(X>1; \mu_0, A_0) =  X^{g_{\alpha}-g_0} \left[\frac{1+A_0 X^{-\mu_0}}{1+A_0}\right].
\ee
Excellent approximations to both $\delta_{\rm B}$ and $f_{\rm B}$ are then given by:
\be
\delta_{\rm B} = \Sigma_{\kappa m} D_{\rm GR}\delta_{i} \equiv \Sigma_{\kappa m} \delta_{\rm GR}, \\
f_{\rm B} = (1+g_{\rm B})f_{\rm GR} = (1+g_{\rm B}) \Omega_{m}^{\gamma_{\rm GR} \approx 0.55}.
\ee
These analytical approximations of both $f_{\rm B}$ and $\delta_{\rm B}$ become exact in the limit $\Omega_{\rm m} \rightarrow 1$.

Using this analytical approximation we can straightforwardly calculate the growth index:
\be
\gamma_{B} \approx \gamma_{\rm GR} + \frac{\ln(1+g_{\rm B})}{\ln \Omega_{\rm m}},
\ee
however since $f_{\rm B}$ does not tend to $1$ as $\Omega_{\rm m} \rightarrow 1$, we follow Ref. \cite{Di Porto:2007ym} in noting that it is arguably better to parametrize $f_{\rm B}$ as $(1+g_{\rm B})\Omega_{\rm m}^{\bar{\gamma}_{\rm B}}$ than as $\Omega_{\rm m}^{\gamma_{\rm B}}$.  In the latter case $\gamma_{\rm B}$ often diverges as $\Omega_{\rm m} \rightarrow 1$ but $f_{\rm B}$ remains finite.  In the former case $\bar{\gamma}_{\rm B} = \gamma_{\rm GR} \approx 0.55$ and $g_{\rm B}$ is as we have calculated.

Figure (\ref{FIGfB}) compares of the exact value (solid red line) of $f_B$, calculated by numerically integrating the perturbation equations, and the analytical approximation (dotted black line) presented above for given values of $\alpha$, $\alpha_0$ and $z_{\ast}$. We also show the GR growth rate as a dot-dashed blue line. We see that in all cases our analytical approximation provides an excellent fit to simulations. We note that because $\alpha > 0$, $f_{B} > f_{\rm GR}$.

\subsection{The Slip Functions: $\Sigma_{\kappa m}$ and $\Sigma_{Im}$}
We noted that if one measures the slip function, $\Sigma$, from either weak-lensing or ISW measurements assuming that the growth rate is unaltered then what is measured is respectively: $\Sigma_{\kappa m}$ and $\Sigma_{Im}$.  In these theories $\Sigma_{\kappa m} = D_{\rm B}/D_{\rm GR}$, and the approximate analytic form of this for a coupling that changes at $z=z^{\ast}$  was given above in Eq. (\ref{SigKapMEqn}).

\FIGURE{\includegraphics[width=7cm]{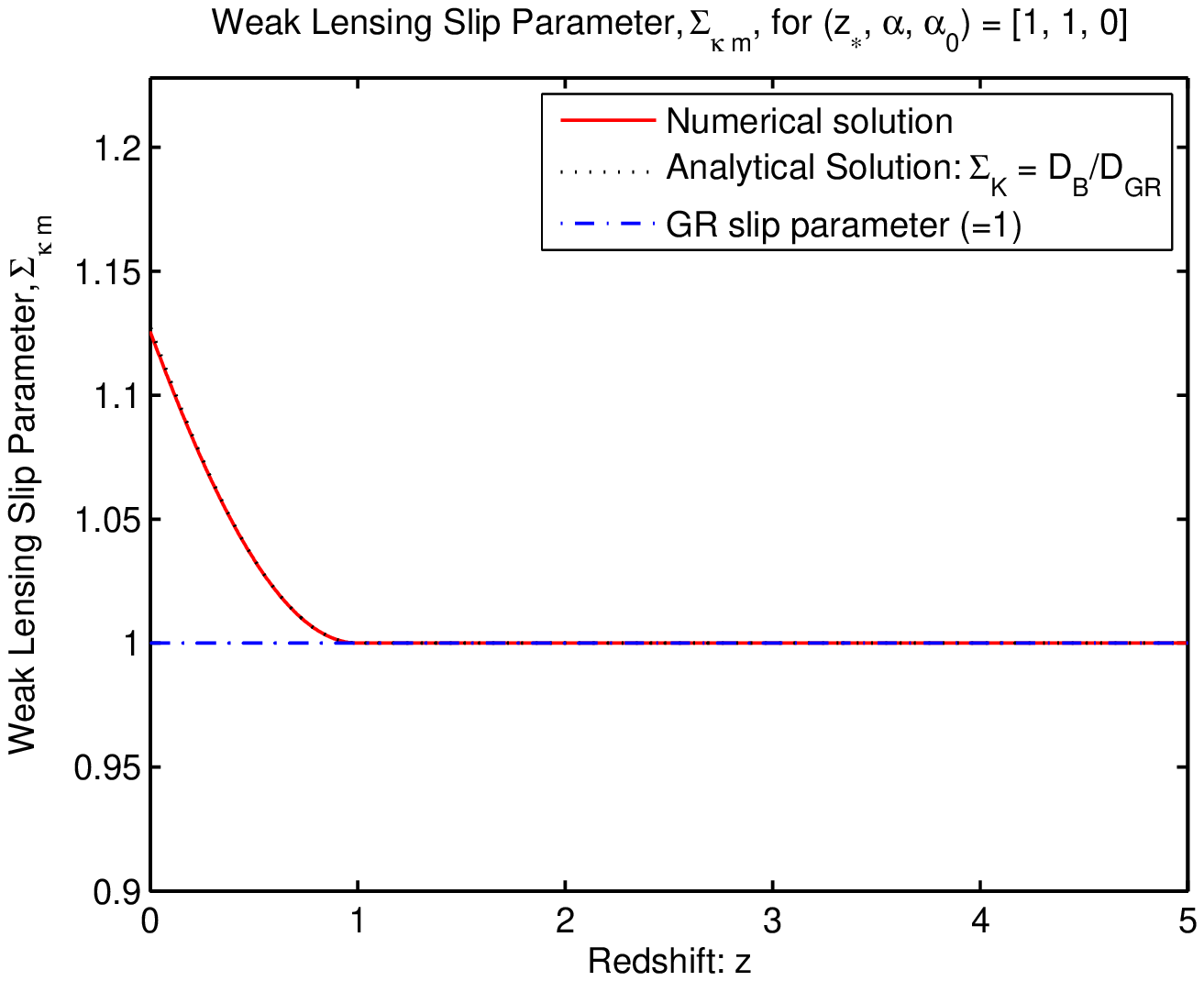}
\includegraphics[width=7cm]{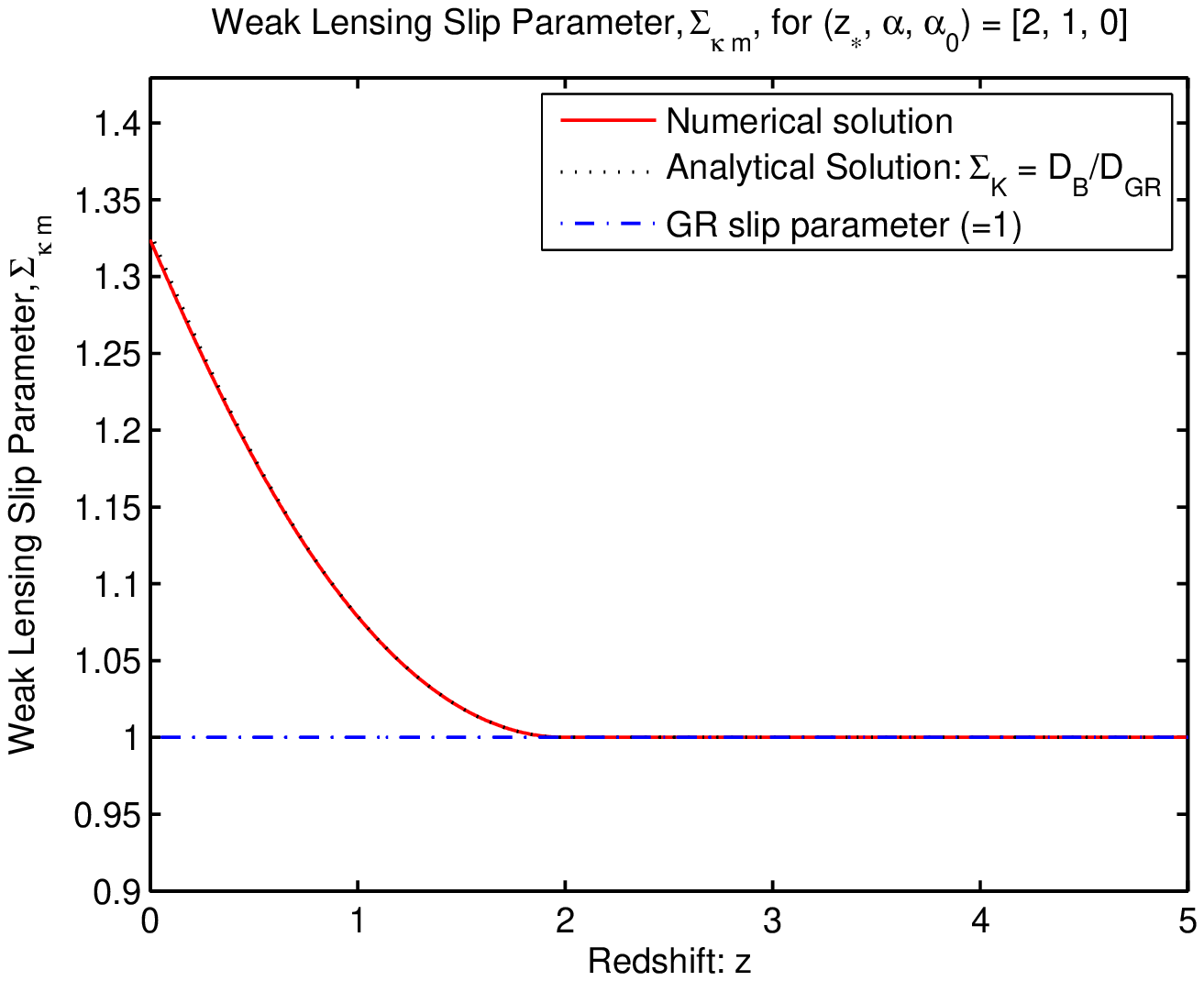}
\includegraphics[width=7cm]{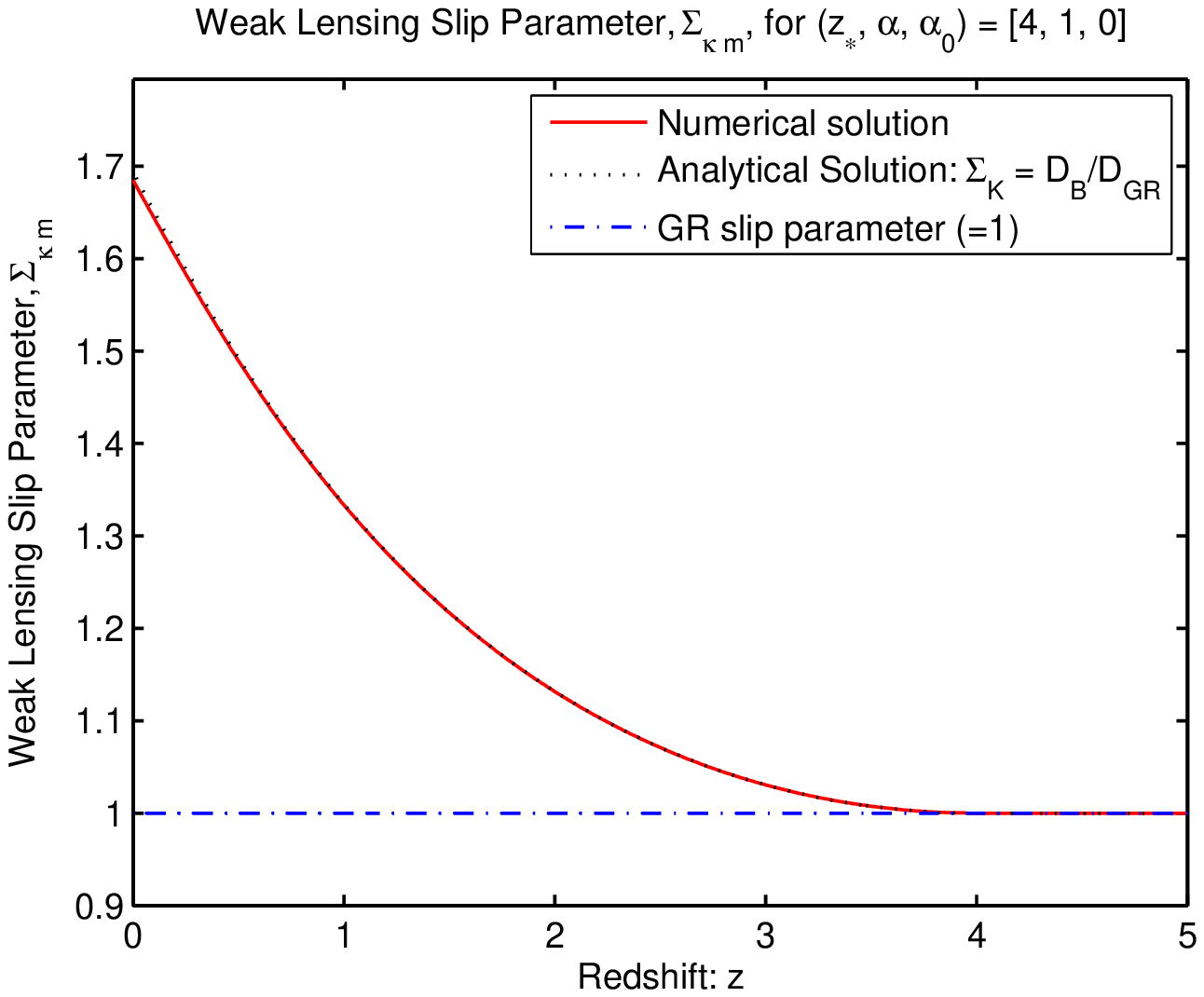}
\includegraphics[width=7cm]{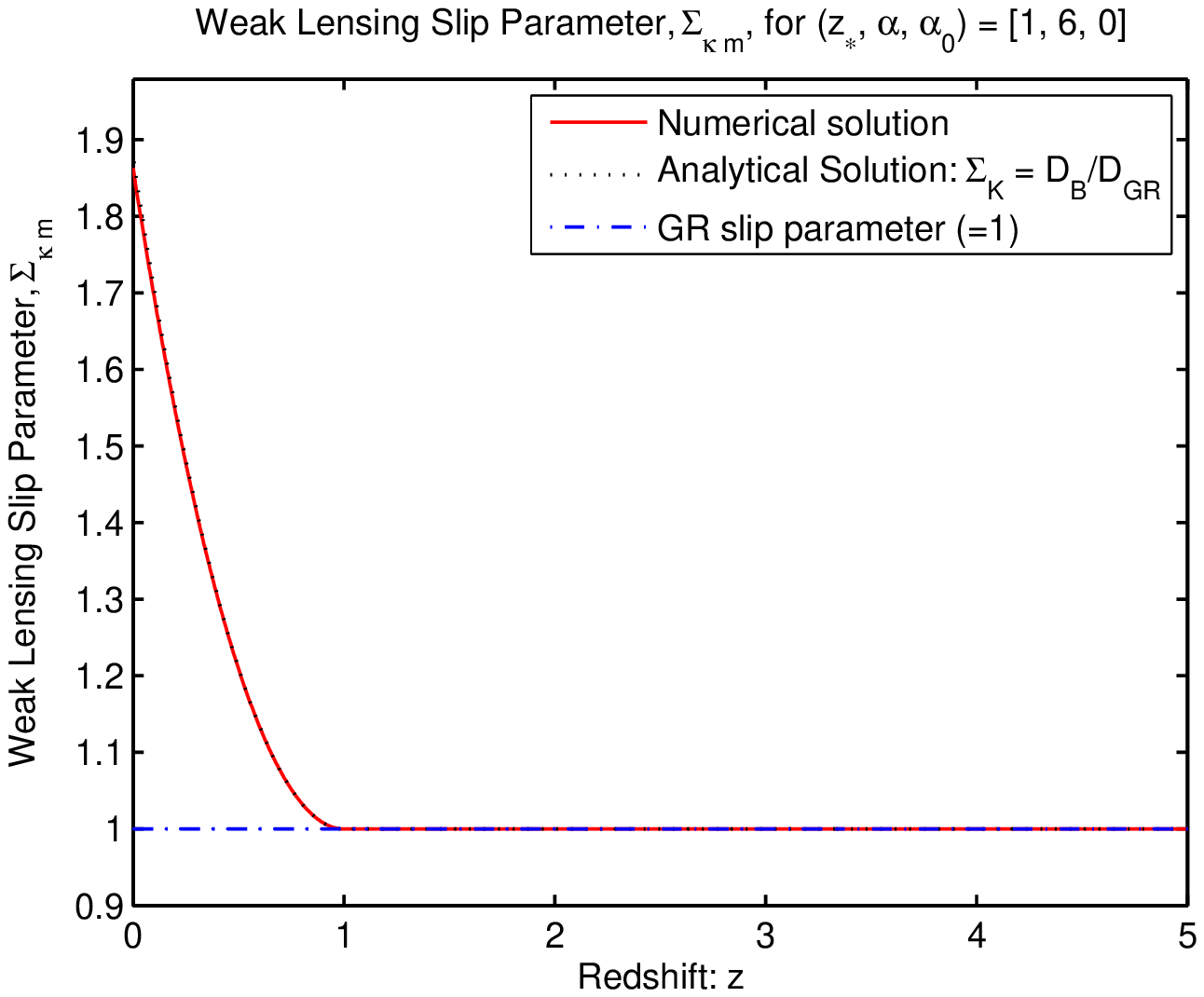}
\caption[]{Sample behaviours for the weak-lensing slip parameter, $\Sigma_{\kappa m} = D_{B}/D_{\rm GR}$, derived under the assumption of a GR growth rate for perturbations. In all cases shown above the $\alpha_0$, fifth force coupling for $z>z_{\ast}$, is taken to vanish and the late-time coupling $\alpha$, for $z < z_{\ast}$, is a constant.  The plots above show the behaviour for $(z_{\ast},\alpha) = (1,1)$ (top-left), $(2,1)$ (top-right), $(4,1)$ (bottom-left) and $(1,6)$ (bottom-right).  In GR, $\Sigma_{\kappa m} = 1$ at all times (shown as the dot-dashed blue line). We see that in all cases, $\Sigma_{\kappa m}$ grows monotonically from an initial value of $1$ for $z< z_{\ast}$.  The solid red line is the exact numerical solution, and the dotted black line is our analytical approximation. We note that the analytical approximation derived in this work is, as with the approximation for the growth-rate, almost exact at all times.}
\label{FIGSigmaK}}

We have also calculated all that is required to have an analytical approximation of the measured ISW slip function $\Sigma_{Im}$ since $\Sigma_{Im}/\Sigma_{\kappa m} = (1-f_{\rm B})/(1-f_{\rm GR})$ and $f_{\rm B} = (1+g_{\rm B})f_{\rm GR}$ so:
\be
\Sigma_{Im}/\Sigma_{\kappa m} = 1-\frac{g_{\rm B}}{f_{\rm GR}^{-1} -1}  \approx 1-\frac{g_{\rm B}}{\Omega_{\rm m}^{-0.55}-1}.
\ee
Now for $\alpha_{\rm BB} >0$, $g_{\rm B}>0$, and since $f_{\rm GR} < 1$ for $\Omega_{\rm m} < 1$ it follows that $\Sigma_{Im}< \Sigma_{\kappa m}$ and it is possible to have $\Sigma_{\kappa m} > 1$ and $\Sigma_{I m} < 1$ at the same time. Additionally, as $\Omega_{\rm m} \rightarrow 1$, $\Sigma_{Im}$ can grow large and negative unless $g_{\rm B}$ decreases at a sufficient rate.  It is feasible that $\Sigma_{\kappa m}$ would show little deviation from its GR value whilst $\vert \Sigma_{I m}-1\vert \sim O(1)$.

\FIGURE{\includegraphics[width=7cm]{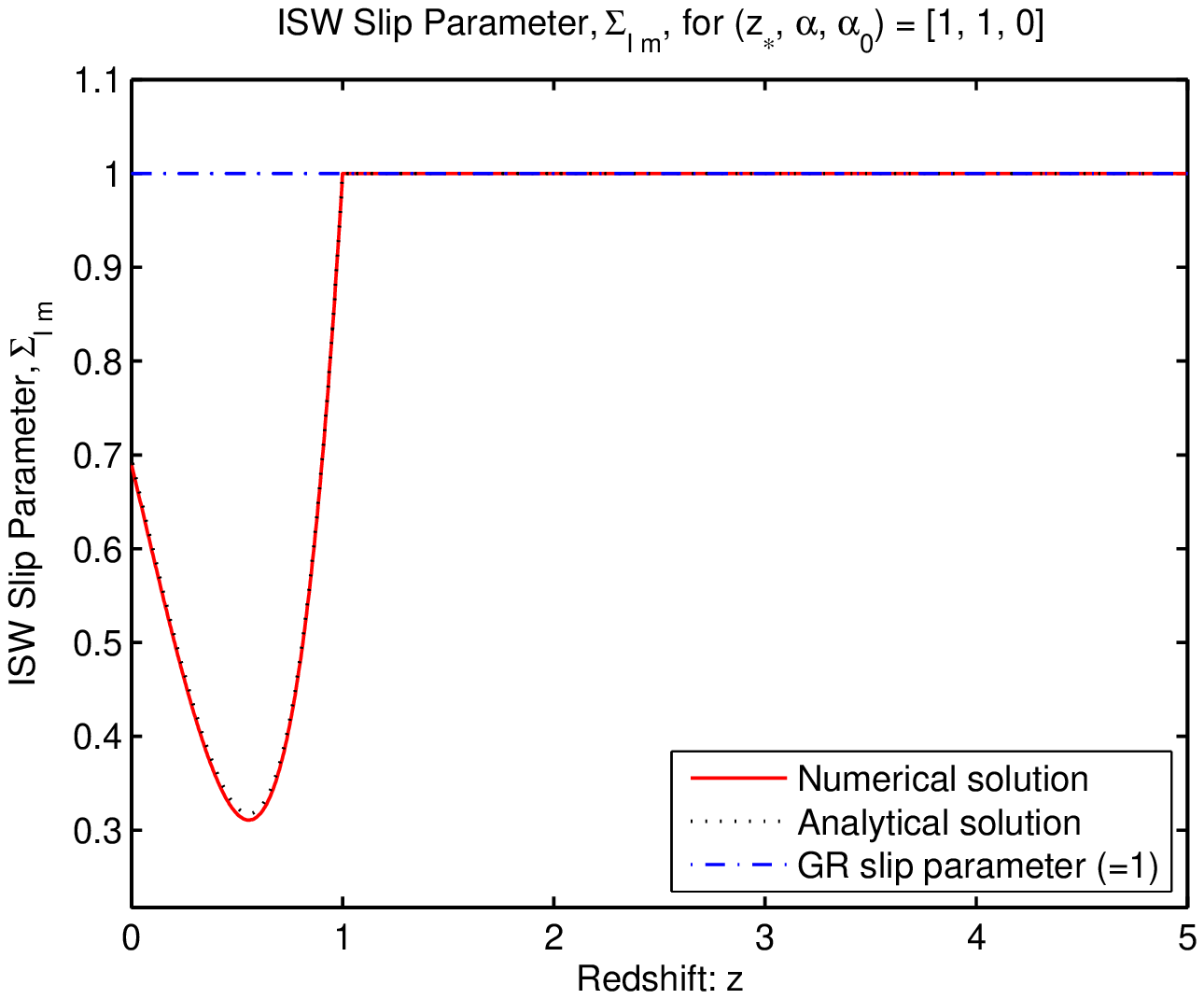}
\includegraphics[width=7cm]{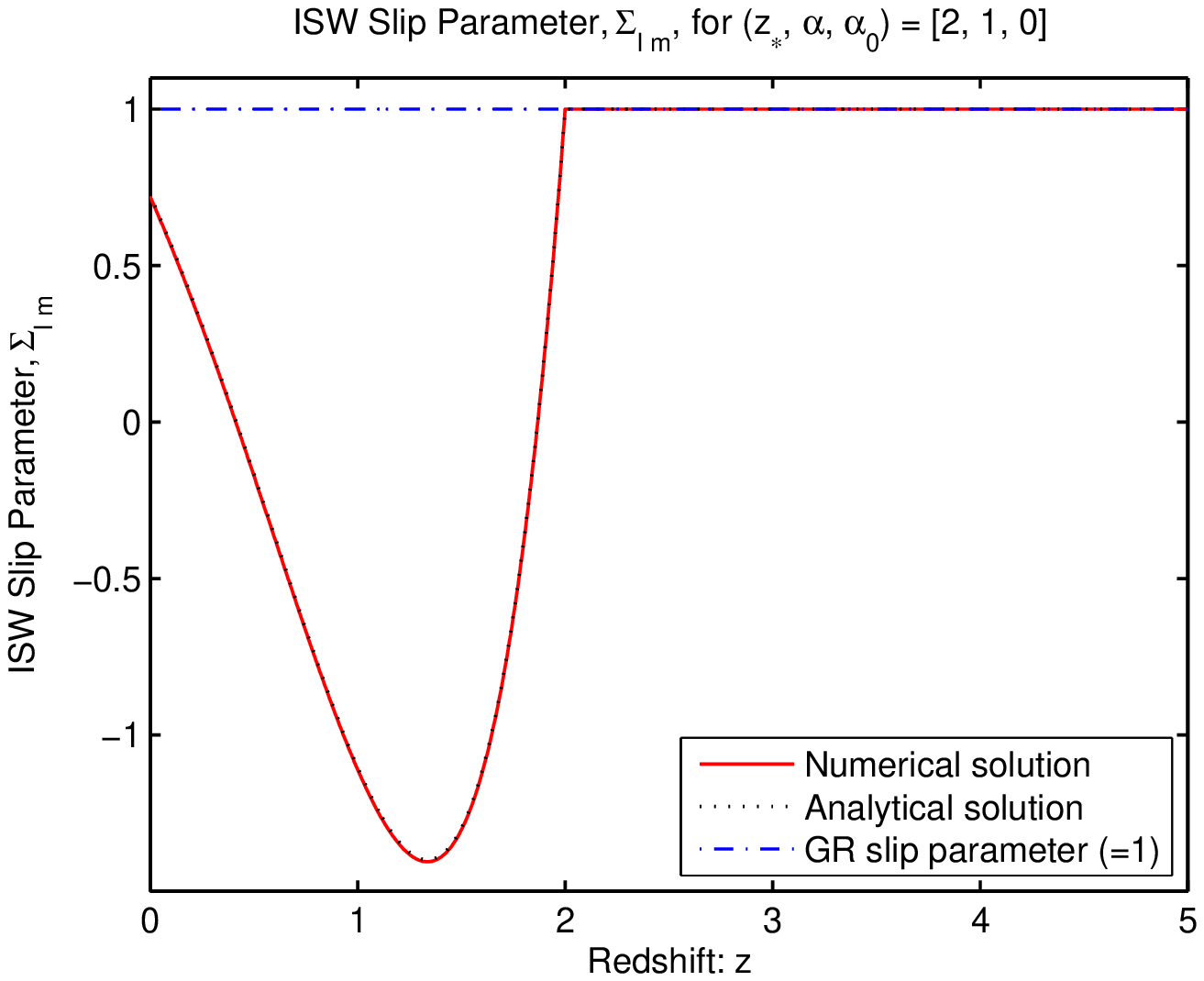}
\includegraphics[width=7cm]{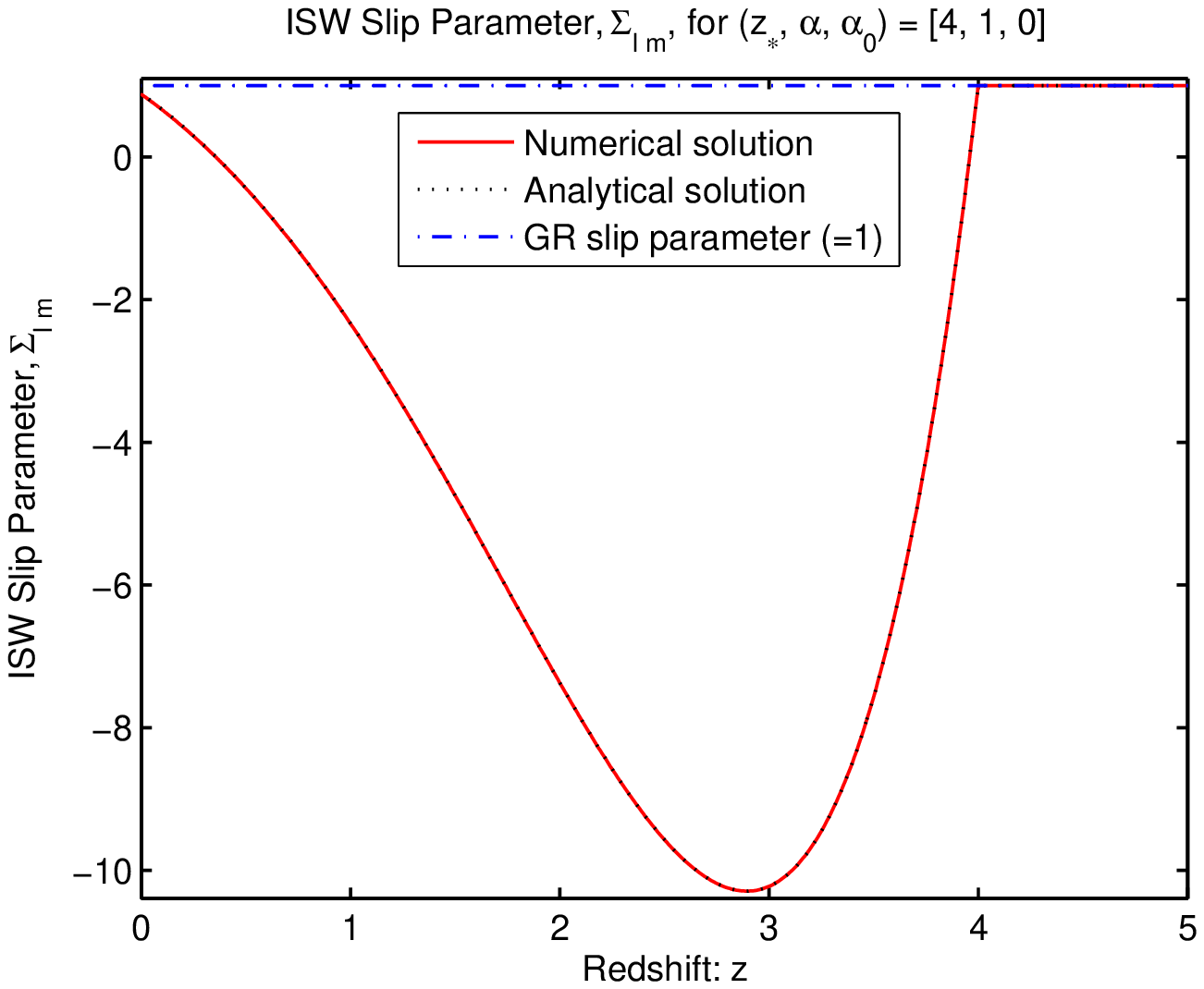}
\includegraphics[width=7cm]{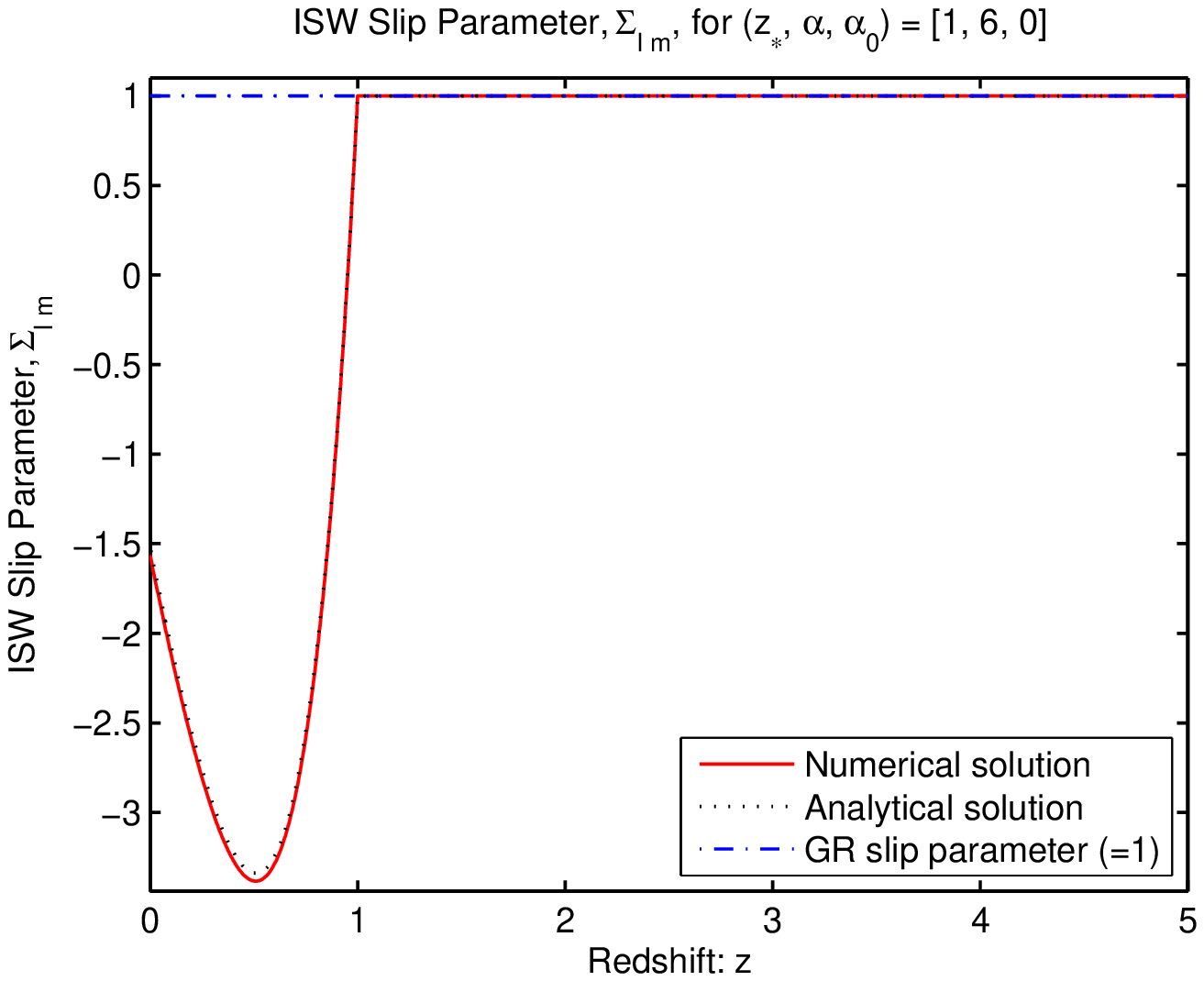}
\caption[]{Sample behaviours for the ISW slip parameter, $\Sigma_{I m}$, derived under the assumption of a GR growth rate for perturbations. In all cases shown above the $\alpha_0$, fifth force coupling for $z>z_{\ast}$, is taken to vanish and the late-time coupling $\alpha$, for $z < z_{\ast}$, is a constant.  The plots above show the behaviour for $(z_{\ast},\alpha) = (1,1)$ (top-left), $(2,1)$ (top-right), $(4,1)$ (bottom-left) and $(1,6)$ (bottom-right).  In GR, $\Sigma_{\kappa m} = 1$ at all times (shown as the dot-dashed blue line). The solid red line is the exact numerical solution, and the dotted black line is our analytical approximation. We note that the analytical approximation derived in this work is, as with the approximation for the growth-rate, almost exact at all times.}
\label{FIGSigmaI}}

Figures (\ref{FIGSigmaK}) and (\ref{FIGSigmaI}) respectively compare of the exact value (solid red line) of $\Sigma_{\kappa m}$ and $\Sigma_{I m}$, calculated by numerically integrating the perturbation equations, with the analytical approximation (dotted black line) presented above for given values of $\alpha$, $\alpha_0$ and $z_{\ast}$. We also show the GR value of $\Sigma_{\kappa m} = \Sigma_{I m} = 1$. In all cases it is clear that our analytical approximation represents an excellent fit to simulations.  We also note that for $\alpha > 0$, $\Sigma_{\kappa m}>1$ and grows monotonically for $z<z_{\ast}$. $\Sigma_{I m}$, however, can be both $>1$ and $<1$, although typically $>1$ values are only found when $z_{\ast}$ is sufficiently in the past.  Additionally $\Sigma_{Im}$ can become negative.  Typically the deviation of $\Sigma_{Im}$ from $1$ is noticeably  more pronounced than that of $\Sigma_{\kappa m}$.

\subsection{Fifth Force Linear Bias: $b_{\chi}$}
We define the quantity $b_{\chi}$ so that
\be
\delta_{A} = b_{\chi}\delta_{B} + \Delta_{0} = b_{AB}\delta_{B}, \nonumber
\ee
where $\Delta_{0}$ is a constant and $b_{AB}$ is the linear bias.  We define the usual linear bias:
$$
b_{\rm lin}(\delta_A) = \left[1 - \frac{\Delta_{0}}{\delta_{A}}\right]^{-1}.
$$
In the absence of a fifth force (i.e. in GR) we then have simply $b_{AB} = b_{\rm lin}$.  In the presence of a fifth force, we have instead
\be
b_{AB}(k,z) = b_{\chi}(k,z) b_{\rm lin}(\delta_A(k,z)), \nonumber
\ee
and so we may see $b_{\chi}$ as being the additional contribution to the linear bias due to the fifth force.

If one measures $\delta_{A}$ then using the form of $b_{\rm lin}$ and the fact that $b_{\rm lin}^{-1}(\delta_A)\delta_A$ is proportional to the initial Gaussian perturbation $\delta_i$, both in GR and this modified gravity scenario, one can measure $\Delta_0$ and hence $b_{\rm lin}$ directly using the bispectrum of $\delta_A$ perturbations. It appears as a shift in the bispectrum of $\delta_A$ compared to the primordial spectrum. It is therefore possible to measure the `bias-corrected' value of $\delta_A$ which is given by $\delta_{A}^{(bc)} = b_{\rm lin}^{-1}\delta_A = b_{\chi}\delta_{B}$.  Thus bias due to $b_{\chi}$ remains even when one corrects for the usual linear bias.

\FIGURE{\includegraphics[width=7cm]{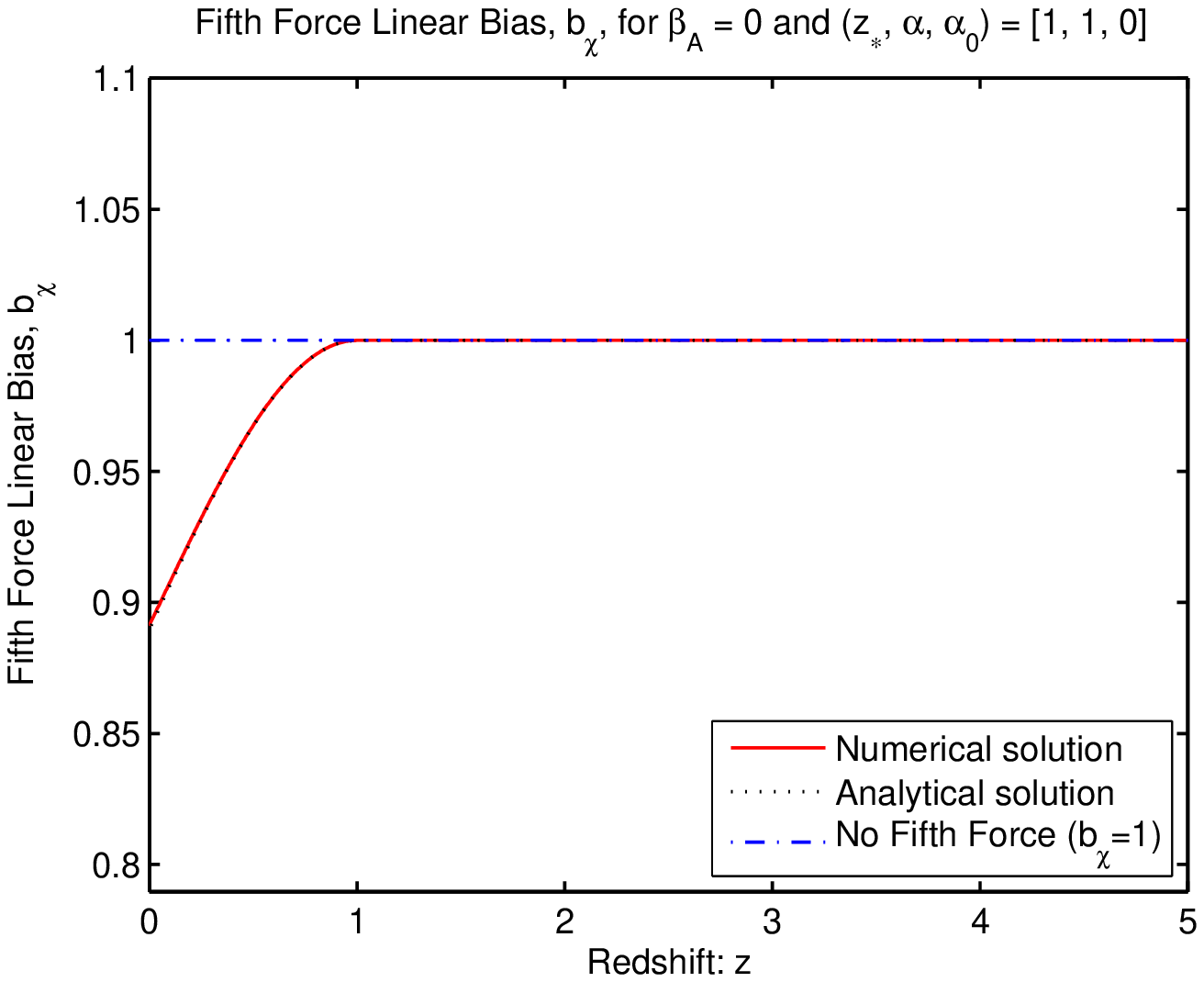}
\includegraphics[width=7cm]{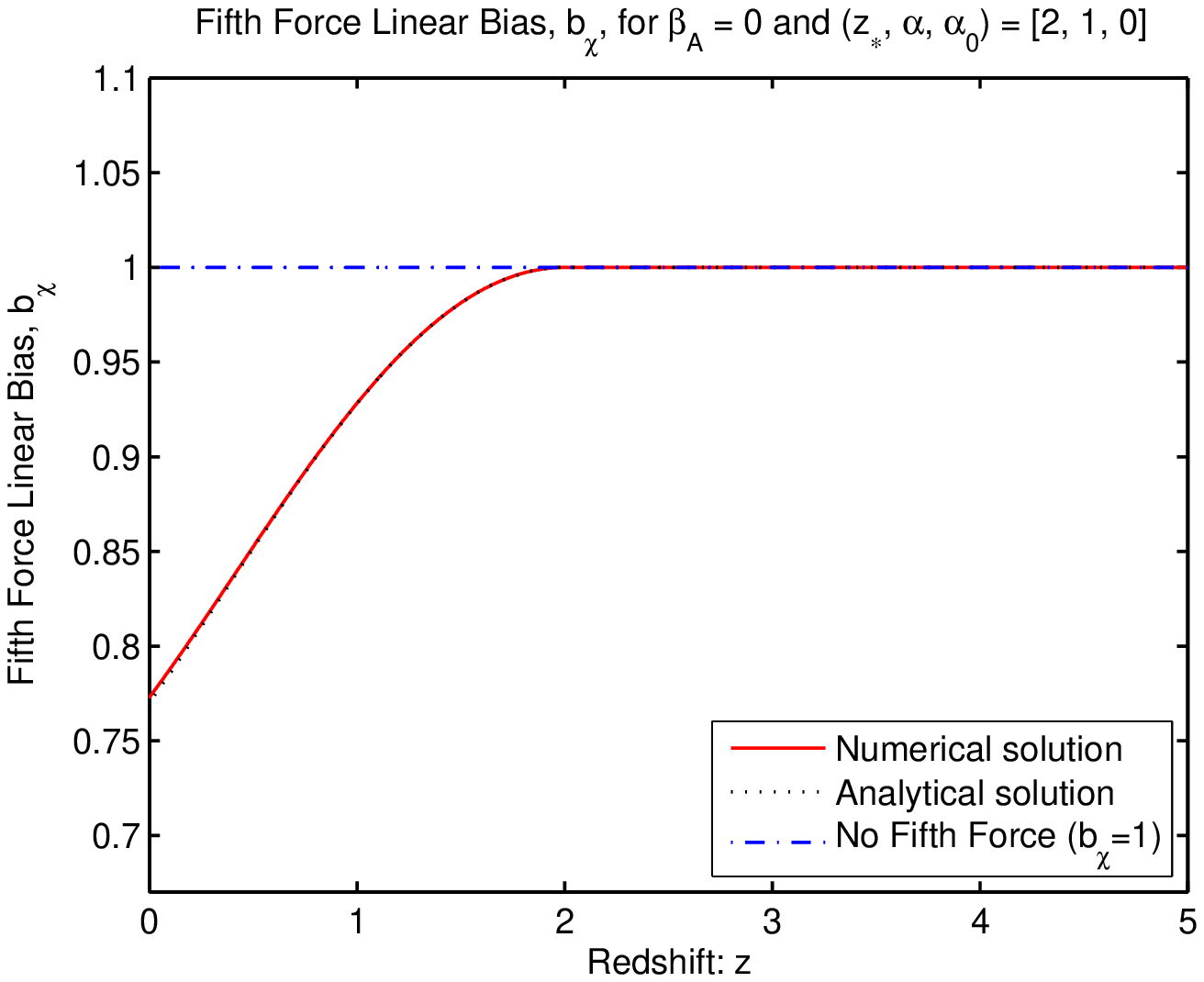}
\includegraphics[width=7cm]{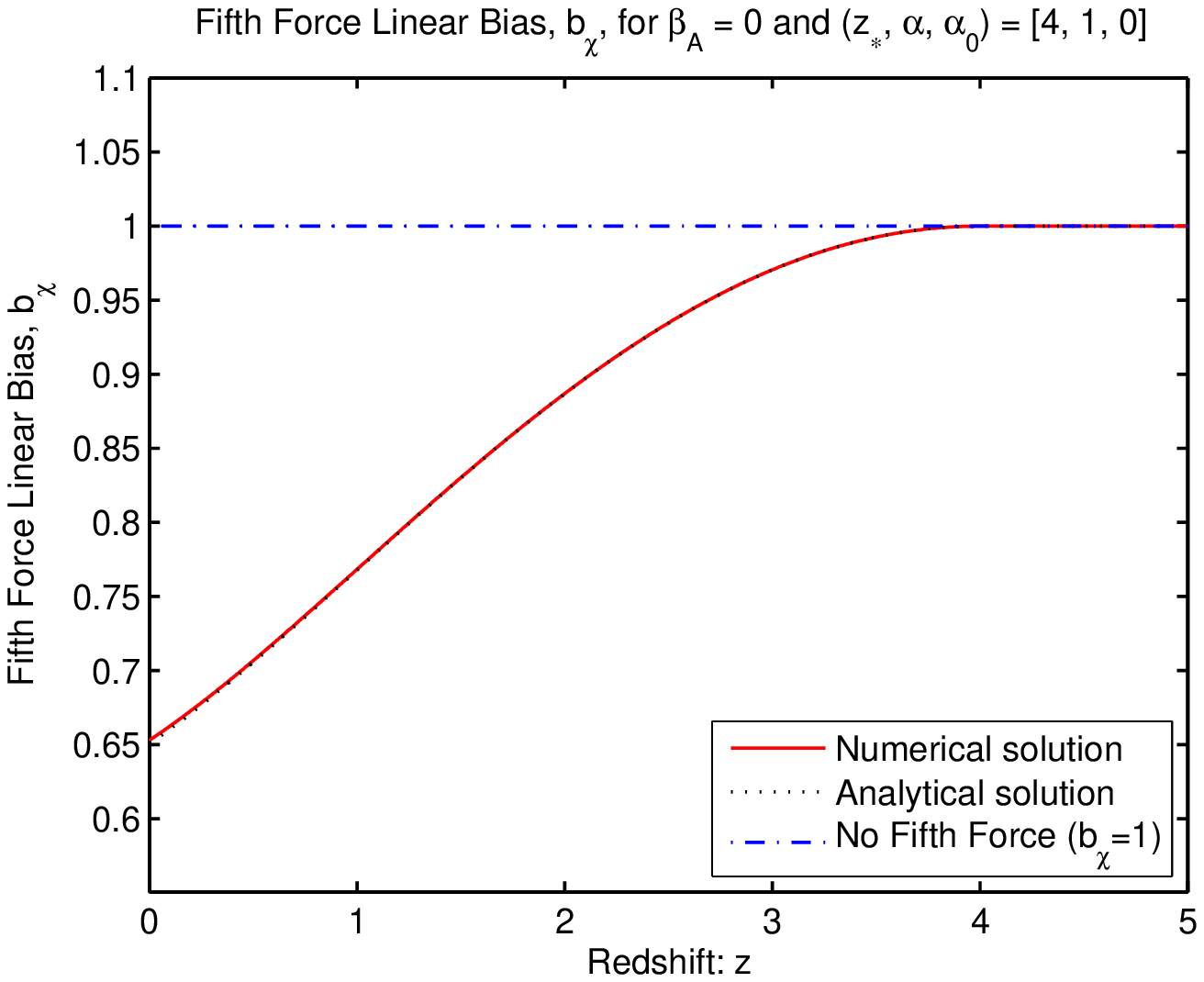}
\includegraphics[width=7cm]{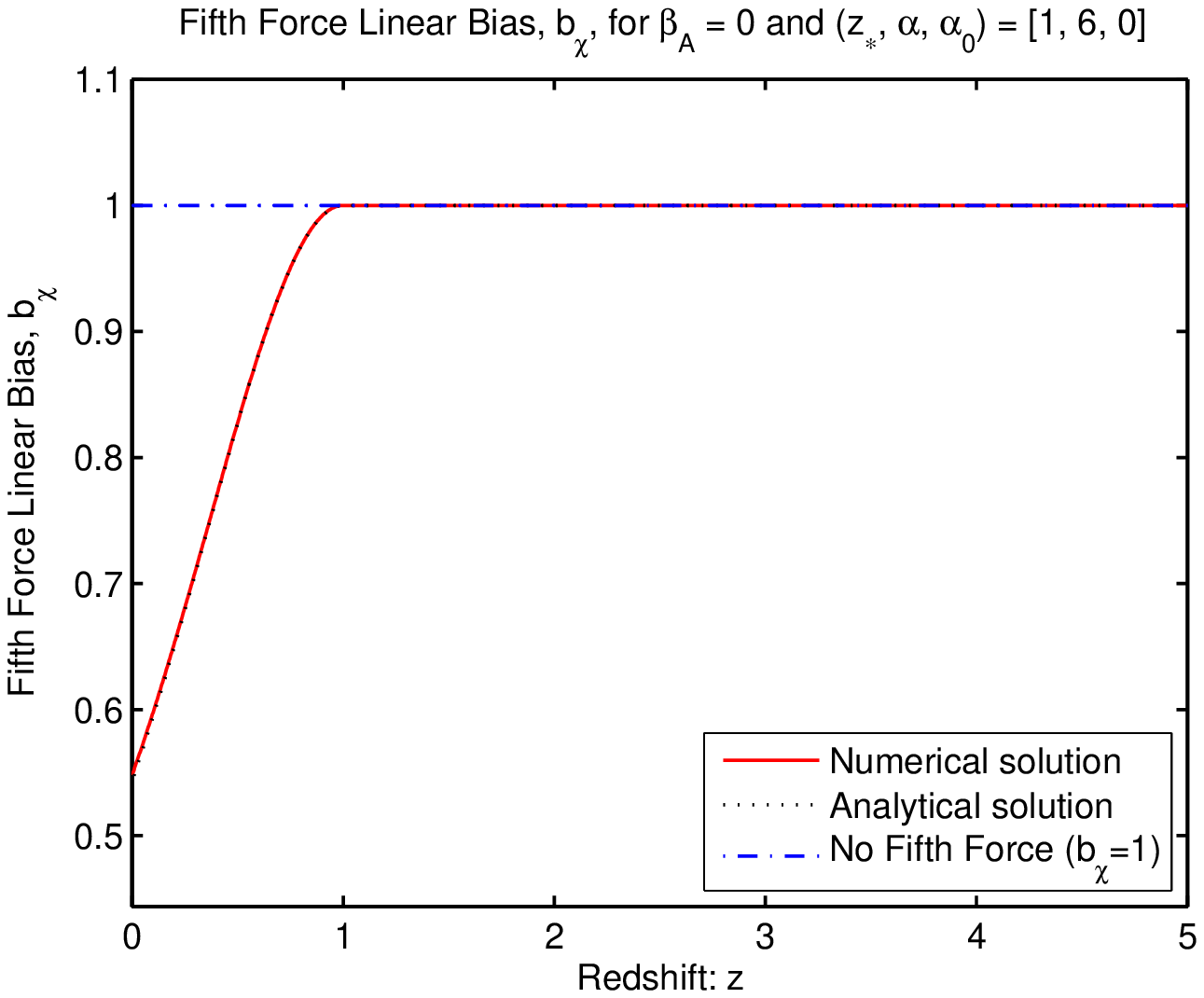}
\caption[]{Sample behaviours for the fifth force linear bias parameter, $b_{\chi} = \delta_A^{(bc)}/\delta_B$. The usual linear bias $b_{\rm lin}$ depends only on $\delta_A$ and can be measured directly using higher statistics of $\delta_A$. In all cases shown above $\alpha_0$, the fifth force coupling for $z>z_{\ast}$, is taken to vanish and the late-time coupling $\alpha$, for $z < z_{\ast}$, is a constant.  The plots above show the behaviour for $(z_{\ast},\alpha) = (1,1)$ (top-left), $(2,1)$ (top-right), $(4,1)$ (bottom-left) and $(1,6)$ (bottom-right).  In GR, $b_{\chi} = 1$ at all times (shown as the dot-dashed blue line). We see that in all cases, $b_{\chi}$ decreases monotonically from an initial value of $1$ for $z< z_{\ast}$.  The solid red line is the exact numerical solution, and the dotted black line is our analytical approximation. We note that the analytical approximation derived in this work is almost exact at all times.}
\label{FIGbchi}}

If species $A$ represents galaxies there will generally be additional multiplicative contributions to the bias relative to the linear CDM perturbation $\delta_{B}$ due to, for instance, the rate of galaxy formation and non-linear effects.   In many viable modified gravity theories, modifications to gravity are suppressed on short-scales and when the ambient density is much larger than the cosmological density. This is the case for chameleon models. So we estimate that the formation and non-linear bias would not be greatly affected by the presence of a cluster scale modification of gravity.   Hence $b_{\chi}$ is therefore expected, in many theories, to represent the dominant additional contribution to the bias due to fifth force effects.

If $\alpha_{\rm AB} = \alpha_{\rm BB}$ at all times (even if $\alpha_{\rm BB} \neq 0$) then $b_{\chi} =1$.  A non-unity value of $b_{\chi}$ only results when species $A$ couples to the fifth force with a different strength than does species $B$.

We now calculate $b_{\chi}$ across a jump at $z=z_{\ast}$.  For $z>z_{\ast}$ we define:
$$
b_{\chi 0} = (1+\alpha_{AB}(z>z_{\ast}))/(1+\alpha_{BB}(z>z_{\ast})), \qquad \alpha_{BB} = \alpha_0,
$$
and for $z<z_{\ast}$,
$$
b_{\chi f} = (1+\alpha_{AB}(z<z_{\ast}))/(1+\alpha_{BB}(z<z_{\ast})), \qquad \alpha_{BB} = \alpha.
$$
For $z>z_{\ast}$, we have simply $b_{\chi} = b_{\chi 0}$.  In the far future of the transition at $z=z_{\ast}$, $b_{\chi} \rightarrow b_{\chi f}$ but generally $b_{\chi}(z<z_{\ast}) \neq b_{\chi f}$.

For $z \leq z_{\ast}$, we define $\Delta b = b_{\chi} -b_{\chi f}$ and then inserting $\delta_{A} = b_{\chi}\delta_B +\Delta_0$ in to the evolution equation for $\delta_A$ we arrive at
\be
Y^{\prime \prime} = \left[\frac{3\Omega_m}{2}-2\right] Y^{\prime}, \label{XEqn}
\ee
where $Y=(\Delta b)\delta_B/\delta_B^{\ast}$.  We define $S = Y^{\prime}/f_{\rm GR} Y$ and using the above identity we have
\be
\frac{S^{\prime}}{f_{\rm GR}} = -\left[\frac{3\Omega_m}{2f_{\rm GR}^2}-1+S\right]S.
\ee
As with the deviation of the $g_{\rm B}$ expression, the only approximation we make is to assume $f_{\rm GR} \approx f_{0} = \Omega_{m}^{-1/2}$.  Then we $3\Omega_m/2f_0^2 -1 =1/2$ and have:
$$
\frac{S^{\prime}}{(\frac{1}{2}+S)S} = -\frac{D_0^{\prime}}{D_0},
$$
and so:
\be
S = \frac{Y^{\prime}}{f_0 Y} \approx \frac{A_1}{2((D_0/D_{\ast})^{1/2}-A_1)}. \label{SEqn}
\ee
Continuity at $z=z_{\ast}$ implies $S_{\ast}=1+g_0$ and so
$$
A_1 = \frac{1+g_0}{\frac{3}{2}+g_0}.
$$
Thus
\be
Y \approx (b_{\chi 0}-b_{\chi f})\left[\frac{1-A_1 (D_0/D_{\ast})^{-1/2}}{1-A_1}\right],
\ee
where we have used $Y_{\ast} =(b_{\chi 0}-b_{\chi f})$.  Using the expression for $\delta_{B}$ derived above and $\mu_0 = 5/2+2g_{\alpha}$  then we obtain
\be
b_{\chi} = b_{\chi 0} + (b_{\chi f} - b_{\chi 0})  \left[1-B(D_0/D_{\ast};g_0,g_{\alpha})\right],
\ee
where
\be
B(X<1;g_0,g_{\alpha}) &=& 1, \\
B(X>1;g_0,g_{\alpha}) &=& X^{-(1+g_{\alpha})}\left[\frac{1+A_0}{1+A_0 X^{-\mu_0}}\right]\left[\frac{1-A_1 X^{-1/2}}{1-A_1}\right].
\ee
This analytical approximation to $b_{\chi}$ is, as with the $\delta_{B}$ approximation, exact when $\Omega_m = 1$ and remains an excellent approximation up to the present era. Figure (\ref{FIGbchi}) compares of the exact value (solid red line) of $b_{\chi}$, found via numerically integration of the  perturbation equations, with the analytical approximation (dotted black line) presented above for given values of $\alpha$, $\alpha_0$ and $z_{\ast}$. We also show the GR value of $b_{\chi} = 1$ as a dot-dashed blue line. In all cases it is clear that our analytical approximation represents an excellent fit to simulations at all times. For $\alpha > 0$, $b_{\chi} < 1$ and decreases monotonically for $z<z_{\ast}$.

\subsection{Growth Function, $f_A$, for Sub-Dominant Matter Species}

It may be the case that one tracks the growth of a perturbation in the dominant matter species, $B$, using a a sub-dominant matter species, $A$.   It would therefore be important to know how the growth function of $\delta_A$ relates  to that of $\delta_B$.

\FIGURE{\includegraphics[width=7cm]{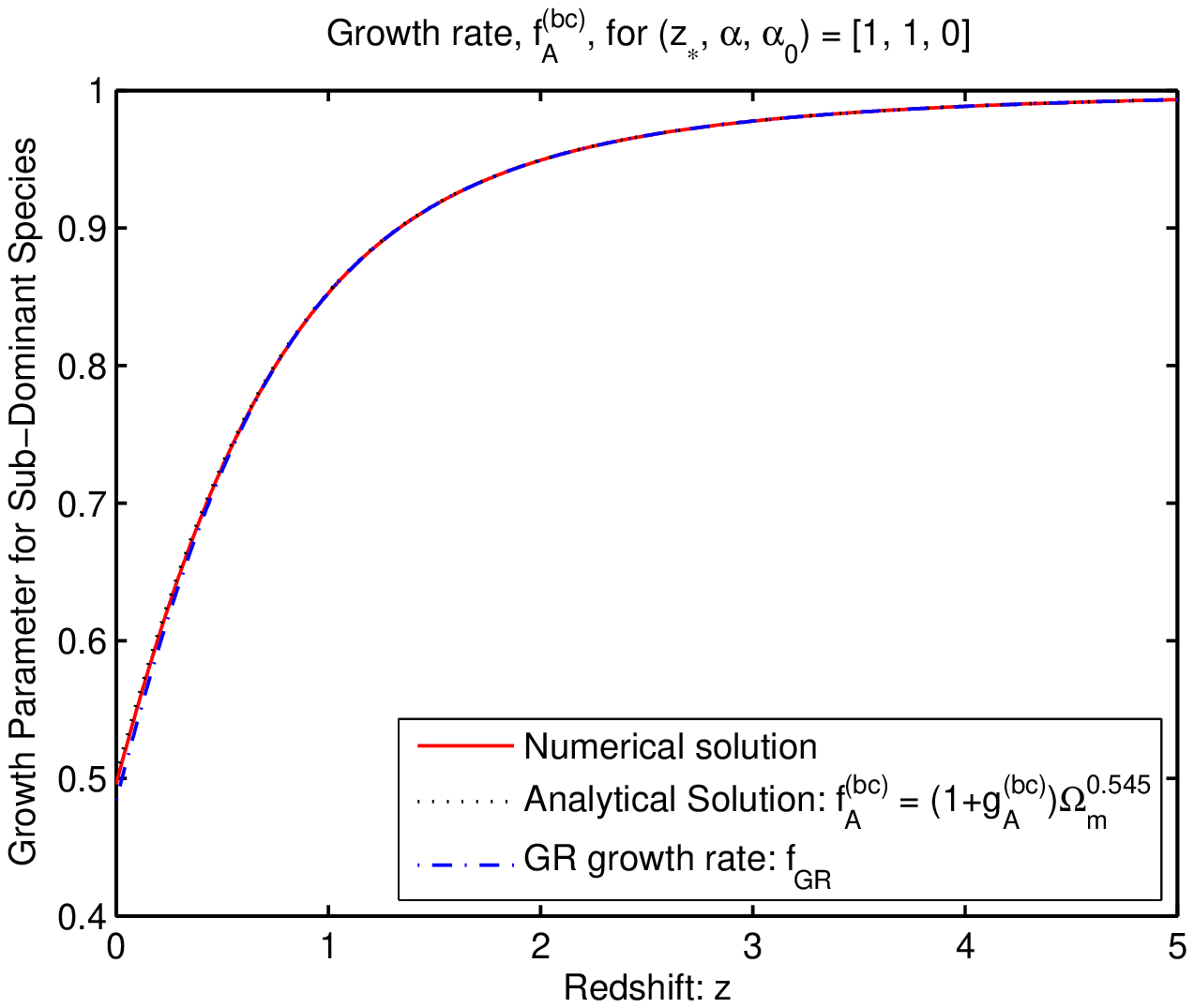}
\includegraphics[width=7cm]{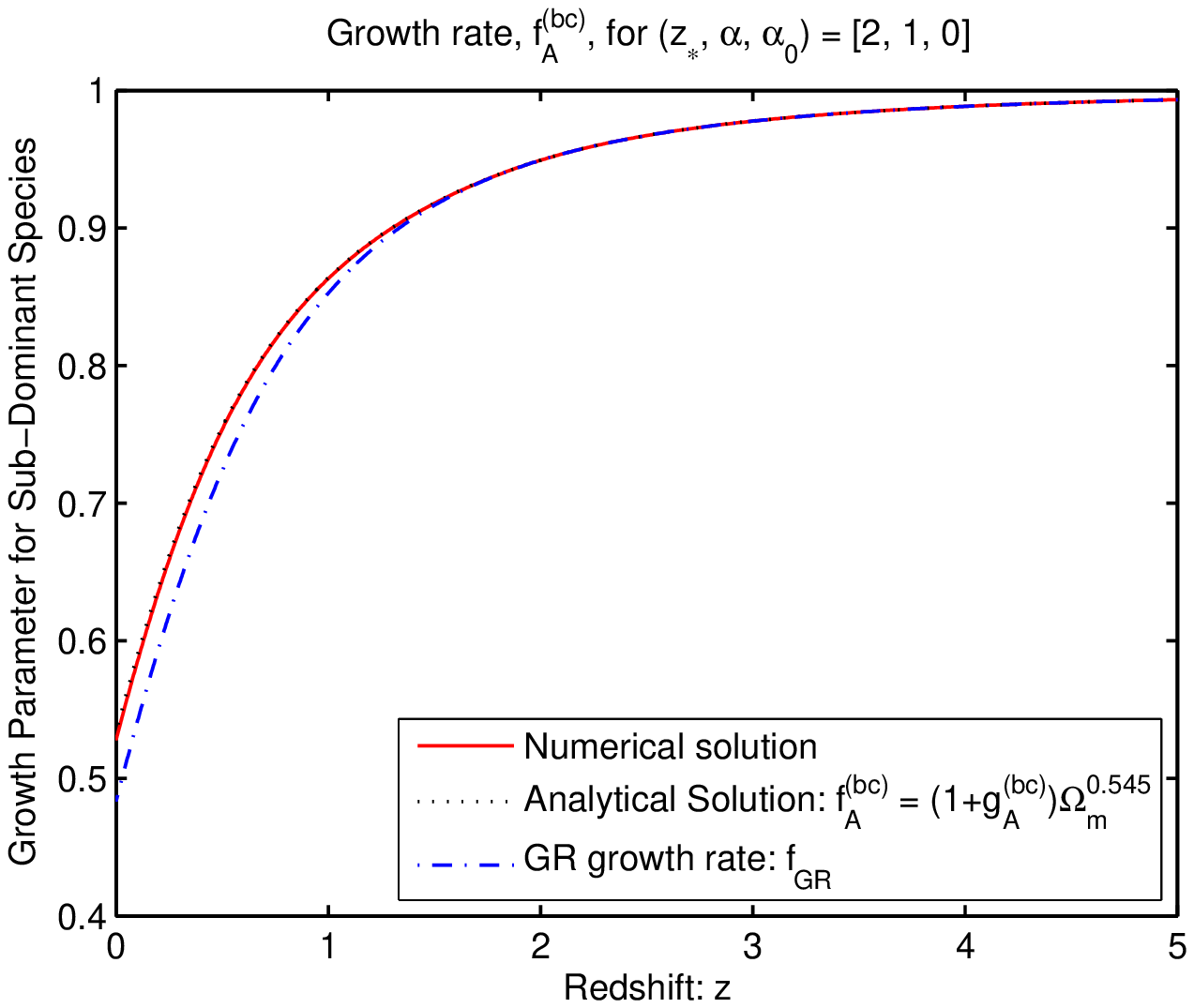}
\includegraphics[width=7cm]{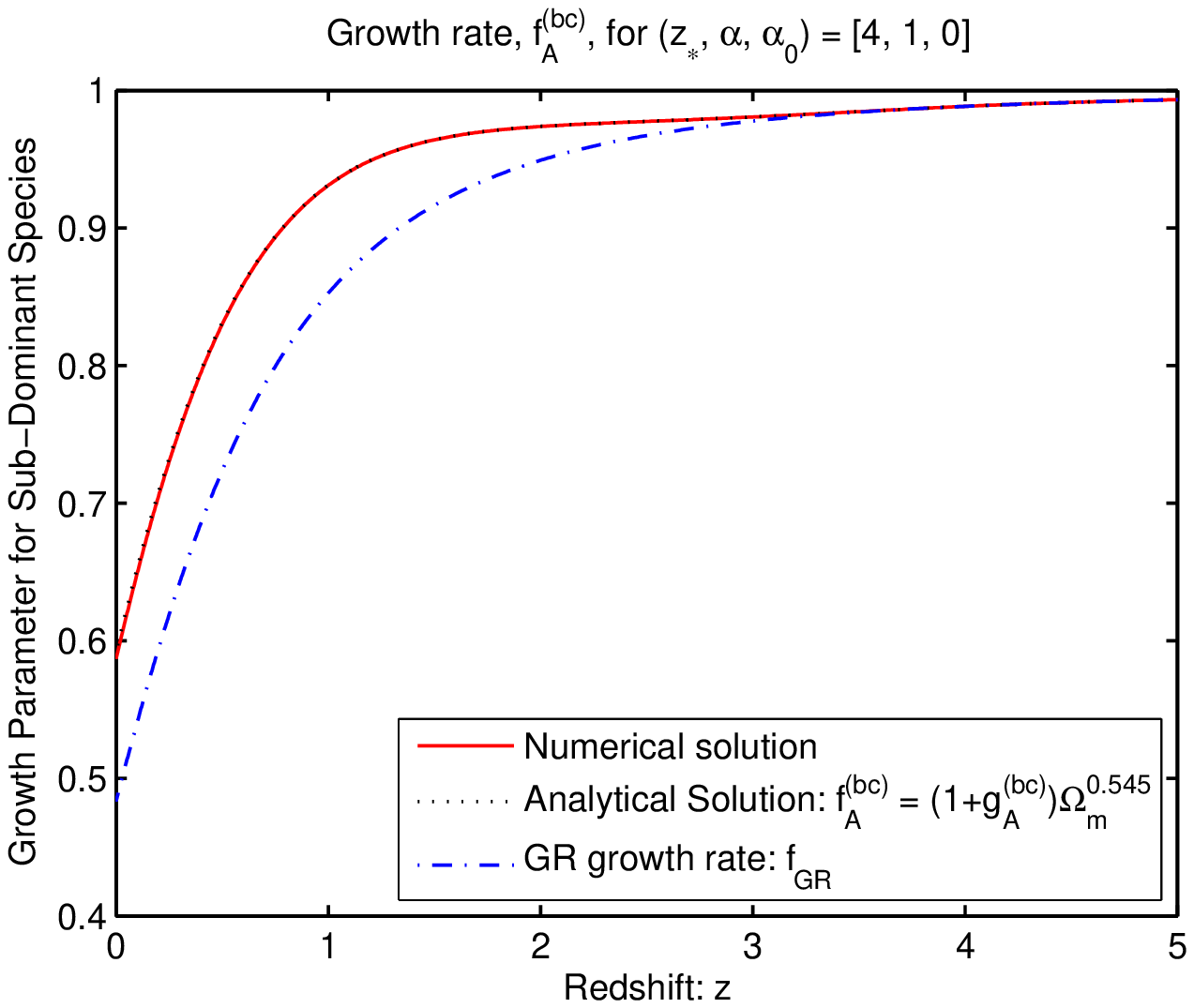}
\includegraphics[width=7cm]{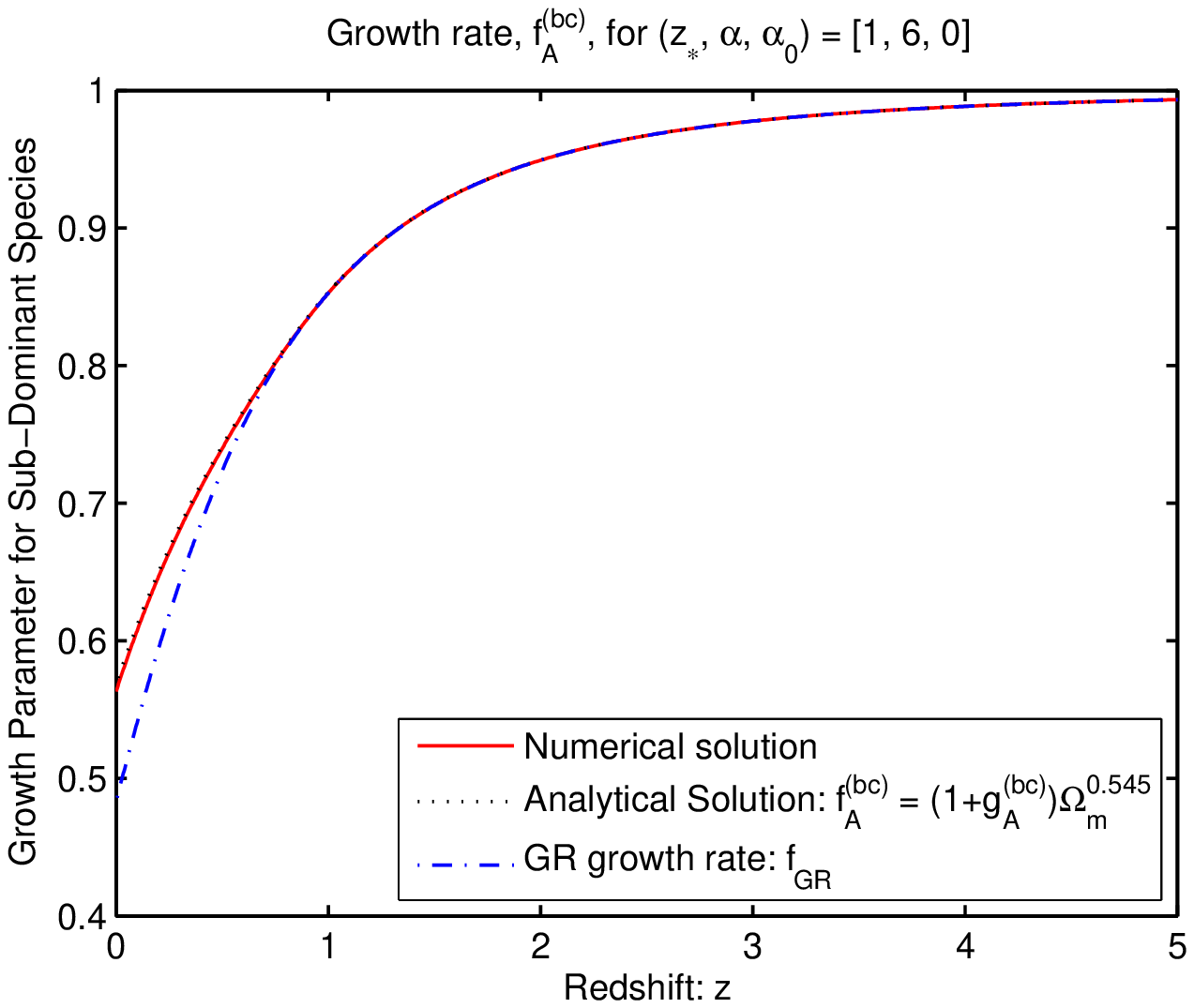}
\caption[]{Sample behaviours for the linear, bias-corrected growth rate, $f_A^{(bc)}$, of a sub-dominant matter species which is uncoupled to the fifth force e.g. galaxies in chameleon / $f(R)$ models. In all cases shown above, $\alpha_0$, the CDM fifth force coupling for $z>z_{\ast}$, is taken to vanish and the late-time CDM coupling $\alpha$, for $z < z_{\ast}$, is a constant.  The plots above show the behaviour for $(z_{\ast},\alpha) = (1,1)$ (top-left), $(2,1)$ (top-right), $(4,1)$ (bottom-left) and $(1,6)$ (bottom-right).   We see that in all cases there is a pronounced deviation from General Relativity where $f_A^{(bc)} = f_{\rm GR} \approx \Omega_{m}^{0.545}$ (shown as the dot-dashed blue line).  The solid red line is the exact numerical solution, and the dotted black line is our analytical approximation: $f_A = (1+g_A)\Omega_{m}^{0.545}$. We note that the analytical approximation derived in this work is almost exact at all times.}
\label{FIGfA}}

Now we have define the `bias-corrected' value of $\delta_A$ to be $\delta_{A}^{(bc)} = b_{\rm lin}^{-1}(\delta_A)\delta_A$; $b_{\rm lin}(\delta_A)$ can be estimated directly from $\delta_A$ measurements using higher order statistics.    Thus whilst estimating bias directly from the data allows one to correct for $b_{\rm lin}$, the $b_{\chi}$ contribution to the bias remains. We therefore consider
\be
\delta_{A}^{(bc)} = b_{\chi}(k,z) \delta_{B}, \label{dAbcEq}
\ee
and define:
\be
f_{A}^{(bc)} = (1+g_{A}^{(bc)})f_{\rm GR} =   \frac{\dd \ln \delta_{A}^{(bc)}}{\dd \ln a}.
\ee
Thus it follows directly from Eq. (\ref{dAbcEq}) that:
\be
g_A^{(bc)} &=&  \left[g_{\rm B}+ \frac{b_{\chi}^{\prime}}{f_{\rm GR}b_{\chi}}\right].
\ee
For $z> z_{\ast}$, $g_{\rm B}=g_0$, $b_{\chi} = b_{\chi 0} = {\rm const}$ and so $g_{A}^{(bc)}(z>z_{\ast}) =g_0$.  For $z<z_{\ast}$ we note that
\be
b_{\chi} = b_{\chi f} + \frac{\delta_{B}^{\ast}}{\delta_{B}}Y,
\ee
where $Y$ obeys Eq. (\ref{XEqn}) and so
\be
\frac{b^{\prime}_{\chi}}{f_{\rm GR}b_{\chi}} = \frac{b_{\chi}-b_{\chi f}}{b_{\chi}} \left[-(1+g_{\rm B}) +\frac{Y^{\prime}}{f_{\rm GR} Y}\right].
\ee
Making the single approximation $f_{\rm GR} \approx f_{0} =\sqrt{\Omega_{m}}$ we have $Y^{\prime}/f_{\rm 0}Y  = S$ where $S$ is given by Eq. (\ref{SEqn}) and so:
\be
g_{A}^{(bc)}(z>z_{\ast}) &=& g_{0} + (g_{\alpha}-g_0) G_{A}(D_0/D_{\ast}), \label{gAa1}
\ee
where
\be
G_A(X<1) &=& 0, \\
G_A(X>1) &=& \frac{b_{\chi f}}{b_{\chi}} \left[\frac{X^{\mu_0}-1}{X^{\mu_0}+A_0}\right] + \frac{b_{\chi f}-b_{\chi}}{b_{\chi}} \left[\frac{1+g_0}{g_{\alpha}-g_0}\right] \left[\frac{X^{1/2}-1}{X^{1/2}-A_1}\right].
\ee

and $f_{A}^{(bc)} = (1+g_A)f_{\rm GR}$.   We can also have $D_{A}^{(bc)} = \delta_{A}^{(bc)}/\delta_{i}$ in terms of $D_0$:
\be
D_{A}^{(bc)} = b_{\chi} \Sigma_{\kappa m} D_{\rm GR}.
\ee
Now when $\beta_A \equiv 0$ i.e. the $A$-type matter species does not feel the fifth force and we have $1+\xi = 1/(1+\alpha)$ and $1+\xi_0 = 1/(1+\alpha_0)$ and $(\xi-\xi_0)/(1+\xi) = -(\alpha-\alpha_0)/(1+\alpha_0)$. Thus, it can be checked that, approximately
\be
\frac{g_A^{(bc)}-g_0}{g_B-g_0} \approx \left(\frac{1+\xi}{1+\xi_{\rm eff}}\right) \frac{x_{\xi}(1+2x_{\xi})}{3}, \label{gAa2}
\ee
where $x_{\xi} = (\xi_{\rm eff}-\xi_0)/(\xi-\xi_0)$; this approximation  is particularly good for small $x_{\xi}$ i.e. close to $z=z_{\ast}$ and at late times when $x_{\xi} \sim 1$.   We also find that (without requiring $\beta_A = 0$), there is a similar approximation for $g_B-g_0$ in terms of $x_{\xi}$. Specifically:
$$
g_B-g_0 \approx (g_{\alpha}-g_0)\tanh\left(\sqrt{\frac{c_0x_{\xi}}{1-x_{\xi}}}\right),
$$
where
$$
c_0 = \frac{3(\alpha-\alpha_0)^2}{(1+\alpha)(g_{\alpha}-g_0)^2}.
$$
Hence we can roughly relate $g_A^{(bc)}$ and hence $\gamma_A$ to $x_{\xi}$ and hence $b_{\chi}$:
\be
\frac{g_A-g_0}{g_{\alpha}-g_0} \approx \left(\frac{1+\xi}{1+\xi_{\rm eff}}\right) \tanh\left(\sqrt{\frac{c_0x_{\xi}}{1-x_{\xi}}}\right)\frac{x_{\xi}(1+2x_{\xi})}{3}. \nonumber
\ee

\noindent The analytical approximation of $g_{A}$ given by Eq. (\ref{gAa1}) is exact at early times and remains very accurate to late times.   Whichever approximation one uses, the growth function, $\gamma_{A}$, is then given by:
\be
\gamma_{A} \approx \gamma_{\rm GR} + \frac{\ln(1+g_A)}{\ln \Omega_{\rm m}}.
\ee
All of the expressions derived have been found using only the approximation $\Omega_m/f_0^2 \sim \Omega_m^{-0.1} \approx 1$ and  hence share the property that when $\Omega_m=1$ they are exact. Fortunately they also remain excellent approximations up to the present day when $\Omega_m \sim 0.26$.   Figure  (\ref{FIGfA}) shows $f_A^{(bc)}$ for  $\beta_A = 0$ at all times for the same values of $\alpha$, $\alpha_0$ and $z_{\ast}$ as were plotted for $f_B$ in figure (\ref{FIGfB}). In each plot the solid red line is the exact numerical solution, the dotted black line is the analytic approximation $f_A^{(bc)} = (1+g_A^{(bc)})\Omega_m^{0.545}$ with $g_A^{(bc)}$ from Eq. (\ref{gAa1}) and the dash-dotted blue line is the GR growth rate, $f_{\rm GR}$.  It is clear that our analytic solution represents an excellent approximation to the exact behaviour of $f_{A}^{(bc)}$.

In all cases, we have considered  $\alpha_0 = 0$ and the coupling to the dominant species (i.e. CDM) turns on at $z=z_{\ast}$. If there were a universal coupling ($\beta_A=\beta_B$) to the fifth force $f_{A}^{(bc)} = f_{B}$ as shown in Figure (\ref{FIGfB}).  By comparing figures (\ref{FIGfB}) and (\ref{FIGfA}), we note that when the coupling to the dominant $B$ matter species  turns on at late times,  the deviation of $f_{A}^{(bc)}$ from its General Relativity value is much less when species $A$ is uncoupled $\beta_A = 0$ than when $\beta_A=\beta_B$.  We note that, whatever value $\beta_A$ takes, $\alpha > 0$, implies $f_A^{(bc)} > f_{\rm GR}$.

We have considered a scenario where the dominant form of matter involved in large scale structure formation i.e. cold dark matter (CDM) on cluster, and larger, scales, is coupled to an additional fifth force mediated by a scalar field, $\chi$. We have denoted such matter to be species $B$ and its  coupling strength $\beta_{B}$ to $\chi$.  The fifth force between species $B$ matter particles is then $\alpha_{BB} = 2\beta_{B}^2$ times the strength of gravity. One does not, however, observe CDM on cluster scales directly, only its effects on observable, non-dark, forms of matter. We have therefore also allowed for a second, subdominant, type of matter dubbed species $A$, whose density contrast
$\delta_A$, and peculiar velocity, $\theta_A$, perturbations can be directly observed and are used to extrapolate information about the density perturbation of large scale distributions of CDM (i.e. species $B$).  Typically, one uses the distribution and velocities of galaxies to measure the distribution of CDM, and so our species $A$ should be taken to be the baryonic matter in the universe (i.e. galaxies).

We have assumed that galaxies, as a whole, have a different (effective) coupling to $\chi$ than CDM.  We denote this coupling by $\beta_{A}$ and defined $\alpha_{AB} = 2\beta_A\beta_B$.  Since galaxies are also predominantly constituted of cold dark matter, in the simplest scalar field models (i.e. those with approximately linear field equations) $\beta_A=\beta_B$, and $\alpha_{AB} = \alpha_{BB}$.  However, there is a key difference between the CDM confined in galaxies and that distributed on large scales, namely the former is much denser than the latter and inside the galaxies the density perturbation compared with the background is highly non-linear.  In scalar field theories with non-linear field equations, a chameleon mechanism might develop which causes the mass of the scalar field to depend on the environment. Typically $\chi$ would then be much heavier in denser regions than it is in sparse regions.  If the mass of the field, $m_{\chi}$, inside galaxies is sufficiently large (i.e. $m_{\chi}^{-1} \ll {\rm few}\,{\rm kpc}$), then the galaxies would effectively decouple from external perturbations in $\chi$ and so there would be almost no fifth force i.e. $\beta_A \ll \beta_B$.  This is realized in both chameleon / chameleonic $f(R)$ theories (in which there is also a coupling to baryonic matter) and the related varying mass dark matter models. It is perfectly feasible therefore that one might find $\alpha_{AB} \ll \alpha_{BB}$ at least when one considers the evolution of perturbations on scales where they are linear i.e. $\delta_B \ll 1$. When the CDM perturbations go non-linear, it is feasible that they might also decouple from the fifth force in the same way as the galaxies have.  This final possibility is beyond the scope of this work and, since it involves complicated non-linear behaviour, mostly likely requires $N$-body simulations to address fully.

\section{Constraints}
In this section we briefly describe the range of possible behaviours for the observables $\gamma_A$ and $\Sigma$ for different values of the couplings $\beta_A$ and $\beta_B$.  We use recent constraints on the growth parameter and $\Sigma$ to give limits on  the different couplings.
 Good limits come from measurements of the galaxy and Lyman-$\alpha$ (Ly$\alpha$) power spectra.  These measure the ratio of the power spectra at one value of $z$ to that at another.  Neither of these tools directly probe the growth rate of the large scale cold dark matter perturbations.  Ly$\alpha$ absorption systems probe the power spectra of the density perturbations in baryons, and galaxy surveys probe the galaxy power spectrum.

The growth rate determined from both galaxy and L$\alpha$ power spectra are usually subject to some bias correction. For the Ly$\alpha$ systems this is generally done by comparing observations with $N$-body simulations which assume GR.  In galaxy surveys such as SDSS the bias is assumed to have a luminosity dependence which is specified a priori up to an overall normalization constant which is fitted for by comparing observations with the CMB linear power spectrum extrapolated to the current epoch (assuming GR growth).  Neither method accounts for the additional red-shift dependent bias, $b_{\chi}$, resulting from a non-universal coupling to the fifth force. We do, however, assume that any bias present in GR has been removed.  Thus we take the measured density perturbation to be $\delta_{A}^{(bc)}$  (up to an overall normalization constant which is degenerate with the average bias).  The measured growth rate is therefore $f_{\rm A}^{(bc)} = (1+g_{A}^{(bc)})f_{\rm GR}$.

Ref. \cite{Di Porto:2007ym} catalogues recent limits on $f_A^{(bc)}$ from Ly$\alpha$ systems and from the 2dFGRS galaxy survey.    The 2dFGRS limit is $f_{\rm gal}^{(bc)}(z=0.15) = 0.49 \pm 0.14$ \cite{Tegmark:2003uf,Guzzo:2008ac}. For this measurement the linear bias was estimated directly from the data essentially by assuming $b_{0}^{-1}(\delta_A)\delta_{A} = D_{A}^{(bc)}(z) \delta_{i}$ where $\delta_i$ is the initial Gaussian fluctuation.   Thus the linear bias estimated in this case will be $b_0(\delta_A)$ in both GR and the class of modified gravity model consider here, and the quoted value of $f$ is truly $f_{A}^{(bc)}$.

Recently Ref. \cite{Guzzo:2008ac} reports another limit of $f_{\rm gal}^{(bc)}$ this time at $z=0.77$ of $f_{\rm gal}^{(bc)}/b_{\rm 0} = 0.70 \pm 0.26$ (VVDS) from the  VIMOS-VLT Deep Survey.  This time, the linear bias, $b_{0}$ was estimated using a different method.  It was assumed that $b_{\rm est} = b_{0}^{{\rm (gal)}}$ where $b_{\rm est} = \sigma_{8}^{(\rm gal)}(z=0.77)/\sigma_{8}^{(\rm cmb)}(z=0.77) = 1.3\pm 0.1$; $\sigma_{8}^{(\rm cmb)\,2}$ is the normalization of the power spectrum amplitude extrapolated from WMAP assuming a GR growth rate.  Thus in this model $b_{\rm est} = b_0^{\rm (gal)} D_{A}^{(bc)}/D_{\rm GR}$.    The constraint on $f$ quoted in Ref. \cite{Guzzo:2008ac} therefore corresponds to $f_{\rm gal}^{(bc)}  D_{A}^{(bc)}/D_{\rm GR} =  0.91 \pm 0.36$ at $z=0.77$.

The limits from Ly$\alpha$ systems are all for $z>2$, and the best constraints are  $\delta_{{\rm Ly}\alpha}^{(bc)}(z=2.72)/\delta_{{\rm Ly}\alpha}^{(bc)}(z=2.125) = 0.83 \pm 0.11$ from Ref. \cite{Viel:2004bf} and $f_{{\rm Ly}\alpha}(z=3) = 1.46 \pm 0.29$ \cite{McDonald:2004xn}.


In what follows we consider the constraints on $\beta_A$, $\beta_B$ and $z_{\ast}$ that follow from the limits on the galaxy / Ly$\alpha$ perturbation growth rate collated.

We consider the $1\sigma$ constraints on $\beta_A$, $\beta_B$ and $z_{\ast}$ that arise from limits on the deviation of $f_{A}^{(bc)}$ and $D_{A}^{(bc)}$ from their GR values. Here species $B$ is the linear (large scale) cold dark matter perturbation, and $A$ is either galaxies or Ly$\alpha$ systems corrected for GR bias.
\TABLE{
\caption[]{The typical 1$\sigma$ constraints on the coupling $\alpha=2 \beta_{\rm CDM}^2(z<z_*)$ of the scalar field to CDM. We have considered two  situations. The first one corresponds to
a universal coupling of the scalar field to all the species $\beta_{\rm gal}=\beta_{\rm CDM}$. In the second case, we assume that clustered objects such as galaxies have no coupling to the scalar field $\beta_{\rm gal}=0$. We have  analysed the role of the coupling of the scalar field using galaxy surveys and Lyman $\alpha$ results. } \label{tableI}
\begin{tabular}{|c|c||c|c|}
\hline
\multicolumn{4}{|c|}{2dFGRS, Ly$\alpha$ and VVDS Data}  \\
\hline \multicolumn{2}{|c||}{Universal Coupling} & \multicolumn{2}{|c|}{$\beta_{\rm gal}=0$}   \\
\hline
\hline $z_{\ast}$ & $\alpha$ & $z_{\ast}$ & $\alpha$  \\
\hline 0.5 & $<$ 0.52 & 0.5 & $<$57.4  \\
\hline 1 & $<$0.31 & 1 & $<$6.7 \\
\hline 2 & $<$0.24 & 2 & $<$1.6 \\
\hline 3 & $<$0.22 & 3 & $<$0.88  \\
\hline 4 & $<$0.35 & 4 & $<$1.3  \\
\hline
\end{tabular}
}
We begin by considering the case of a universal coupling when $\beta_A=\beta_B = \beta$, that turns on at some redshift $z=z^{\ast}$.  So $2\beta^2(z>z_{\ast}) = 0$ and $2\beta^2(z<z_{\ast}) = \alpha$.  We assume a flat $\Lambda$CDM background with $\Omega_{\rm m} = 0.27\pm0.1$.   The 1-$\sigma$ limits on $\alpha$ for different $z^{\ast}$ from the 2dFGRS, VVDS and Ly$\alpha$ data are shown in left half of table \ref{tableI}.  These are derived by  finding the $\alpha$ that minimizes the $\chi^2(\alpha)$ for the data \cite{DiPorto:2007ym}; $\bar{\alpha}$ say.    The 1-$\sigma$ confidence limit corresponds to those value of $\alpha$ for which $\chi^2(\alpha) -\chi^2(\bar{\alpha})<1$.     For $z_{\ast}>3$, the constraints on $\alpha$ weaken slightly because of large value of $f_{{\rm Ly}\alpha}$ at $z=3$ extrapolated from the Ly$\alpha$ data.

Next we consider the limits on $\alpha=2\beta_B^2(z<z_{\ast})$ under the assumption that galaxies effectively do not couple to the fifth force $\beta_{\rm gal} = 0$.  We assume that $\beta_{{\rm Ly}\alpha} = \beta_{B}$.  Again for $z>z_{\ast}$, $\beta_B= 0$ and the coupling turns on at $z=z_{\ast}$.  Such a scenario would be realized, for instance, in $f(R)$ and chameleon models where the effective coupling is density dependent and as such virialized objects such a galaxies would be uncoupled.   These limits are given in table \ref{tableI}.   Table \ref{tableIb} shows limits on $\alpha$ for scenarios where the fifth forces couples only to dark matter and baryons are uncoupled.  In this case $\beta_{{\rm Ly}\alpha} = 0$ but $\beta_{\rm gal} \approx \beta_{B}$ (since galaxies are predominantly dark matter).  Also in table \ref{tableIb}, we display growth rate limits on $\alpha = 2\beta_{B}^2$ when both galaxies and Lyman $\alpha$ fluctuations are uncoupled, $\beta_{\rm gal}=\beta_{{\rm Ly}\alpha}=0$.  This would be realized, for instance, in dark sector chameleon models where the fifth forces affects only cold dark matter and the effective coupling is density dependent.  When $\beta_{{\rm Ly}\alpha} = 0$, the constraints are very similar to those for $\beta_{{\rm Ly}\alpha} = \beta_{B}$ for $z_{\ast} \lesssim 3$ but become stronger for larger $z_{\ast}$.  This is primarily due to the large observational value of $f_{{\rm Ly}\alpha}(z=3)$  which, alone, slightly prefers $\beta_{{\rm Ly}\alpha} > 0$.

\TABLE{
\caption[]{The typical 1$\sigma$ constraints on the coupling $\alpha=2 \beta_{\rm CDM}^2(z<z_*)$ of the scalar field to CDM. In both situations displayed here the coupling to baryons and hence Lyman$\alpha$ fluctuations is taken to vanish ($\beta_{{\rm Ly}\alpha}=0$). We have then considered two  situations. The first one corresponds to
a universal coupling of the scalar field to all dark matter $\beta_{\rm gal}=\beta_{\rm CDM}$. In the second case, we assume that clustered objects such as galaxies have no coupling to the scalar field $\beta_{\rm gal}=0$. We have  analysed the role of the coupling of the scalar field using galaxy surveys and Lyman $\alpha$ results. } \label{tableIb}
\begin{tabular}{|c|c||c|c|}
\hline
\multicolumn{4}{|c|}{2dFGRS, Ly$\alpha$ and VVDS Data}  \\
\hline \multicolumn{2}{|c||}{$\beta_{{\rm Ly}\alpha} = 0$} & \multicolumn{2}{|c|}{$\beta_{{\rm Ly}\alpha} = \beta_{\rm gal}=0$}  \\
\hline
\hline $z_{\ast}$ & $\alpha$ & $z_{\ast}$ & $\alpha$  \\
\hline 0.5 & $<$ 0.52 & 0.5 & $<$57.4  \\
\hline 1 & $<$0.31 & 1 & $<$6.7  \\
\hline 2 & $<$0.24 & 2 & $<$1.6  \\
\hline 3 & $<$0.22 & 3 & $<$0.87 \\
\hline 4 & $<$0.22 & 4 & $<$0.65 \\
\hline
\end{tabular}
}

\section{Concluding Remarks}

In this paper we considered a scalar-tensor theory with one scalar degree of freedom, whose coupling to matter is not  universal. The couplings of cold dark matter, baryons, neutrinos, etc. to the scalar field is not the same, which will affect the growth of perturbations in the different matter species.

To study large scale structure formation in this setup we considered two non-relativistic fluids: a dominant spe\-cies $B$ (playing the role of cold dark matter) and a subdominant species $A$ (representing baryonic matter). We have found analytic formulae for the growth rate of the perturbations in both species if the couplings are constant. Furthermore we considered the case of perturbations crossing the Compton wave\-length of the scalar field. In doing so, we assumed that the (effective) coupling of the scalar field to the different matter species changed suddenly in the past. We were able to find approximate formulae for the growth rate.

We then discussed the phenomenology of the theory and showed that  there are new observational signatures predicted by the theory such as an anomalous growth of structures influenced by a time dependent growth index $\gamma$ for both Cold Dark Matter and baryons.  Most importantly, there is  a single slip parameter $\Sigma_\kappa$ relating the two Newton potentials, but several ways of extracting this  slip parameter from data. Our analysis provides a template for the growth of structures in the context of a well-motivated theoretical framework, i.e. scalar-tensor theories where the scalar couples to different matter species with  constant couplings.  Different cosmological experiments (such those which probe the ISW, weak lensing or the distribution of galaxies) would probe the slip parameter and the growth of structures, and therefore the forces in the dark sector. It would be interesting to study how well future planned cosmological experiments can probe the type of effects discussed in this paper.

\section{Acknowledgements}
One of us, DJS, is supported by STFC.
The work of CvdB and ACD is supported in part by STFC.

\end{document}